\pdfoutput=1
\documentclass[sigconf,nonacm,screen]{acmart}

\usepackage{amsmath}

\usepackage{orcidlink} 
\usepackage{amssymb}
\usepackage{amsthm}
\usepackage{mathtools}
\usepackage{bm}              
\usepackage{bbm}             
\usepackage{braket}          
\usepackage{physics}         
\usepackage{booktabs}
\usepackage{tabularx}
\usepackage{array}
\usepackage{makecell}   
\usepackage{multirow}
\usepackage{colortbl}
\usepackage{graphicx}
\usepackage{xcolor}
\usepackage{hyperref}
\usepackage{cleveref}
\usepackage{algorithm}
\usepackage{algpseudocode}
\usepackage{enumitem}
\usepackage{microtype}
\usepackage{float}
\usepackage{placeins}
\usepackage{acro}
\newcommand{\mc}[1]{\mathcal{#1}}
\newcolumntype{Y}{>{\raggedright\arraybackslash}X}
\newtheorem{definition}{Definition}

\setcitestyle{numbers,sort&compress}
\AtBeginDocument{%
  }

\newcommand{\StudyExp}{\textsuperscript{\textsf{\bfseries E}}}
\newcommand{\StudyNum}{\textsuperscript{\textsf{\bfseries N}}}

\DeclareAcronym{BPTT}{
short = BPTT,
long = backpropagation through time
}

\DeclareAcronym{CPTP}{
short = CPTP,
long = completely positive trace preserving
}

\DeclareAcronym{CV}{
short = CV,
long = continuous variable
}

\DeclareAcronym{ESN}{
short = ESN,
long = echo state network
}

\DeclareAcronym{ESP}{
short = ESP,
long = echo state property
}

\DeclareAcronym{ENSO}{
short = ENSO,
long = El Ni{\~n}o Southern Oscillation
}

\DeclareAcronym{FID}{
short = FID,
long = free induction decay
}

\DeclareAcronym{FMP}{
short = FMP,
long = fading memory property
}

\DeclareAcronym{FGSM}{
short = FGSM,
long = fast gradient sign method
}

\DeclareAcronym{GBS}{
short = GBS,
long = Gaussian boson sampling
}

\DeclareAcronym{GPU}{
short = GPU,
long = graphics processing unit
}

\DeclareAcronym{GRU}{
short = GRU,
long = gated recurrent unit
}

\DeclareAcronym{HPC}{
short = HPC,
long = high performance computing
}

\DeclareAcronym{IPC}{
short = IPC,
long = information processing capacity
}

\DeclareAcronym{LSTM}{
short = LSTM,
long = long short term memory
}

\DeclareAcronym{MC}{
short = MC,
long = memory capacity
}

\DeclareAcronym{MSE}{
short = MSE,
long = mean squared error
}

\DeclareAcronym{MPI}{
short = MPI,
long = message passing interface
}

\DeclareAcronym{NALM}{
short = NALM,
long = nonlinear amplifying loop mirror
}

\DeclareAcronym{NARMA}{
short = NARMA,
long = nonlinear autoregressive moving average
}

\DeclareAcronym{NISQ}{
short = NISQ,
long = noisy intermediate scale quantum
}

\DeclareAcronym{NMR}{
short = NMR,
long = nuclear magnetic resonance
}

\DeclareAcronym{NVAR}{
short = NVAR,
long = nonlinear vector autoregression
}

\DeclareAcronym{PCA}{
short = PCA,
long = principal component analysis
}

\DeclareAcronym{PGD}{
short = PGD,
long = projected gradient descent
}

\DeclareAcronym{QELM}{
short = QELM,
long = quantum extreme learning machine
}

\DeclareAcronym{QML}{
short = QML,
long = quantum machine learning
}

\DeclareAcronym{QLSTM}{
short = QLSTM,
long = quantum long short term memory
}

\DeclareAcronym{QPU}{
short = QPU,
long = quantum processing unit
}

\DeclareAcronym{QRC}{
short = QRC,
long = quantum reservoir computing
}

\DeclareAcronym{RC}{
short = RC,
long = reservoir computing
}

\DeclareAcronym{RL}{
short = RL,
long = reinforcement learning
}

\DeclareAcronym{RNN}{
short = RNN,
long = recurrent neural network
}

\DeclareAcronym{SSM}{
short = SSM,
long = state space model
}

\DeclareAcronym{VQC}{
short = VQC,
long = variational quantum circuit
}

\begin{document}

\ccsdesc[500]{Computing methodologies~Machine learning approaches}

\title{Quantum Reservoir Computing: Recent Advances and Future Directions}

\author{Shehbaz Tariq}
\authornote{These authors contributed equally to this research.}
\email{shehbaz.tariq@uni.lu}
\orcid{0000-0003-2701-2725}
\affiliation{%
  \institution{Interdisciplinary Centre for Security, Reliability and Trust
  (SnT), University of Luxembourg}
  \city{Luxembourg City}
  \country{Luxembourg}
}

\author{Muhammad Talha}
\authornotemark[1]
\email{muhammad.talha@uni.lu}
\orcid{0009-0002-4793-6650}
\affiliation{%
  \institution{Interdisciplinary Centre for Security, Reliability and Trust
  (SnT), University of Luxembourg}
  \city{Luxembourg City}
  \country{Luxembourg}
}

\author{Symeon Chatzinotas}
\email{symeon.chatzinotas@uni.lu}
\orcid{0000-0001-5122-0001}
\affiliation{%
  \institution{Interdisciplinary Centre for Security, Reliability and Trust
  (SnT), University of Luxembourg}
  \city{Luxembourg City}
  \country{Luxembourg}
}

\author{Arshid Ali}
\authornotemark[1]
\email{Arshid.Ali@tees.ac.uk}
\orcid{0000-0002-7813-846X}
\affiliation{%
  \institution{School of Computing, Engineering and Digital Technologies,
  Teesside University}
  \city{Middlesbrough}
  \country{United Kingdom}
}

\author{Muhammad Diyan}
\email{m.diyan@tees.ac.uk}
\orcid{0009-0003-7337-4708}
\affiliation{%
  \institution{School of Computing, Engineering and Digital Technologies,
  Teesside University}
  \city{Middlesbrough}
  \country{United Kingdom}
}

\renewcommand{\shortauthors}{Tariq et al.}

\begin{abstract}
Quantum reservoir computing (QRC) uses the dynamics of a fixed or weakly tuned quantum system to transform temporal and sequential inputs into measured features, while training is typically confined to a classical readout. This separation reduces reliance on repeated quantum parameter updates and avoids the barren plateaus associated with variational circuit training. Its computational power is often attributed to the exponentially large Hilbert space of the quantum system. However, the memory, nonlinearity, and expressivity that determine what a reservoir can actually compute depend jointly on the input encoding, quantum evolution, observables, measurement, and readout, not on Hilbert space dimension alone. On hardware these capabilities are further constrained by finite sampling, hardware noise, measurement backaction, and the cost of estimating observables, so a large state space alone does not guarantee useful computation. In this survey, we develop a common system model that connects these components and use it to organize QRC foundations, computational properties, reservoir architectures, operating protocols, and physical implementations. We examine spin, photonic, superconducting, bosonic, neutral atom, and other analog platforms, together with applications, software and high performance computing support, benchmarking, and reproducibility. The analysis distinguishes hardware demonstrations from simulations and identifies the assumptions and resources that govern comparisons across implementations. Current results do not establish a broad quantum advantage over well matched classical reservoirs. We therefore specify the resource accounting, benchmark standards, and theoretical criteria needed to evaluate claims of quantum advantage.
\end{abstract}
\acresetall


\keywords{Quantum Reservoir Computing, Quantum Machine Learning, Echo State
Networks, Quantum Advantage, Noisy Intermediate-Scale Quantum Computing, Temporal Information Processing, Physical Reservoir Computing, Quantum Extreme Learning Machines}

\maketitle 

\section{Introduction}
\label{sec:introduction}

\subsection{Background}
Learning from a sequence requires a model to preserve information that remains
useful for the task while discarding details from the distant past. This
requirement arises in chaotic signal forecasting, sequential pattern
recognition, and other time dependent problems
\cite{ZV:23:IEEE_O_ACC}. Feedforward networks map each supplied input to an
output without carrying an internal state between successive samples, so
temporal context must be encoded in the input itself
\cite{ZV:23:IEEE_O_ACC,V:24:MTVEAIIMP56SC}. In contrast, \acp{RNN} propagate an
evolving hidden state, but \ac{BPTT} can encounter vanishing or exploding
gradients during training \cite{PMB:ND:P3ICICMLV2}. These difficulties motivate
learning paradigms that retain dynamical memory without optimizing all
recurrent parameters \cite{V:24:MTVEAIIMP56SC}.

\Ac{RC} addresses this optimization bottleneck by mapping an input sequence through recurrent dynamics whose internal parameters remain unchanged, while training only the readout, which is commonly linear \cite{V:24:MTVEAIIMP56SC,TYRNKEtAl:19:NN,YHBTLEtAl:24:NC}. The resulting state provides a high dimensional feature representation in which temporal information and nonlinear transformations of the input can be accessed by the readout \cite{V:24:MTVEAIIMP56SC,MANBGEtAl:21:AQT}. For a fixed reservoir feature matrix, readout training often reduces to regularized linear regression and therefore avoids backpropagation through the recurrent dynamics \cite{TYRNKEtAl:19:NN,YHBTLEtAl:24:NC}. This distinction is especially useful in physical reservoirs because modifying internal dynamics can require hardware control or calibration \cite{V:24:MTVEAIIMP56SC,TYRNKEtAl:19:NN}. Some recent designs tune a small set of internal control parameters while retaining a simple readout, providing an intermediate regime between conventional reservoirs and fully trained recurrent networks \cite{SPG:25:NC}.

An analogous optimization bottleneck appears in \ac{QML}. On current \ac{NISQ} hardware, one common approach uses \acp{VQC}, whose parameters are trained through repeated quantum executions and classical updates. This process can require many circuit evaluations and measurements and can encounter barren plateaus \cite{CABBEEtAl:21:NRP_2,ACCCC:21:Qu,SL:25:Qu}. Quantum kernel methods provide an alternative based on quantum feature mapping and kernel estimation. Fixed kernels avoid circuit parameter training, whereas trainable kernels reintroduce an optimization loop; both require state preparation and measurements \cite{E:25:IEEE_J_TQE}. At scale, these approaches fit hybrid workflows in which a \ac{QPU} generates quantum outputs while classical \ac{HPC} resources handle preprocessing, optimization, and workflow control \cite{SGNMWEtAl:26:FGCS}. Quantum reservoir architectures follow this hybrid allocation by using quantum dynamics for temporal feature generation and classical resources for training and orchestration \cite{FN:17:PRA_2,SPKRCEtAl:24:NC}.

\Ac{QRC} typically leaves the reservoir dynamics fixed and trains a classical readout over measured observables \cite{FN:17:PRA_2,FN:20:arxiv,MANBGEtAl:21:AQT}. Implementations span spin networks, neutral atom arrays, nonlinear oscillators, and gate based circuits \cite{FN:17:PRA_2,AHZWWEtAl:24:arxiv,GRRKO:21:PRR,CNY:20:PRA_2}. The motivation is that the reservoir state evolves in a space that grows exponentially with system size: for an \(N\) qubit reservoir the Hilbert space dimension is \(2^N\), which suggests a high dimensional state representation at modest hardware size. However, only features exposed by the input encoding, dynamics, and measurements can contribute to prediction \cite{FN:17:PRA_2,TTNLBEtAl:25:arxiv}, so this nominal dimension overstates the usable feature space. Nonlinear and temporal representations therefore depend on the encoding, evolution, observable set, measurement protocol, and readout \cite{GRRKO:21:PRR,TN:20:arxiv,TTNLBEtAl:25:arxiv}. Current results are task specific and do not establish a broad advantage over well matched classical reservoirs \cite{AHZWWEtAl:24:arxiv,TTNLBEtAl:25:arxiv}.

\subsection{Survey Scope and Related Surveys}
\label{sec:methodology-related-surveys}

The survey is organized around a \ac{QRC} processing chain: input preprocessing,
quantum encoding, reservoir dynamics, measurement, readout, and execution
resources. This organization permits systems implemented on different physical
substrates to be compared by computational role and makes clear when similar
hardware is operated with different encoding or measurement protocols. The
literature is grouped by foundations, computational properties, architectures,
operating protocols, hardware, applications, software and \ac{HPC} workflows,
and evaluation practice. General \ac{RC}, \ac{QML}, and \ac{HPC} studies are
included only when they supply transferable definitions, classical baselines,
or resource accounting needed to interpret \ac{QRC} results.

Prior \ac{QRC} reviews and perspectives explain the main formulations, quantum
substrates, measurement considerations, and representative tasks
\cite{MANBGEtAl:21:AQT,FN:20:arxiv,GWPSSEtAl:25:arxiv}. Reviews of physical and
photonic \ac{RC} provide broader accounts of substrate classes, input and output
interfaces, and application settings
\cite{TYRNKEtAl:19:NN,ASAPSEtAl:25:JPP,AGM:24:Dy}. General \ac{RC} surveys cover
theory, algorithms, software tools, physical implementations, and applications,
while dedicated benchmark reviews examine task design and evaluation practice
\cite{YHBTLEtAl:24:NC,ZV:23:IEEE_O_ACC,WTS:25:IJPEDS}. These works have
complementary aims and organizing principles, but no single prior survey
connects all stages of a \ac{QRC} implementation with software and \ac{HPC}
workflows, resource accounting, and reproducible comparison.

Table~\ref{tab:survey_comparison} summarizes this topical coverage. The symbols
indicate whether a topic receives sustained, partial, or no substantive
treatment; they do not assess quality or impact. The contribution of this survey
is to connect these topics through one system model and \ac{QRC} taxonomy. This
structure links architectures and protocols to hardware constraints,
applications, software and \ac{HPC} workflows, resource accounting, and
benchmark requirements. The following subsection states the resulting
contributions in detail.

\begin{table*}[ht]
\centering
\caption{Topic coverage in prior \acs{QRC}, physical \acs{RC}, and general \acs{RC} surveys and perspectives relative to the scope of this survey.}
\label{tab:survey_comparison}
\renewcommand{\arraystretch}{1.3}
\setlength{\tabcolsep}{2pt}
\footnotesize
\begin{tabularx}{\textwidth}{@{}>{\centering\arraybackslash}m{0.105\textwidth}>{\raggedright\arraybackslash\hspace{0pt}}X*{7}{>{\centering\arraybackslash}m{0.042\textwidth}}@{}}
\toprule
\rowcolor{black!5}
\textbf{Scope} &
\makecell[l]{\textbf{Related work}\\\textbf{and focus}} &
\textbf{Fnd.} &
\textbf{Tax.} &
\textbf{P/R} &
\textbf{HW} &
\textbf{Tasks} &
\textbf{SW} &
\textbf{Eval.} \\
\midrule
\multirow{3}{0.105\textwidth}{\centering\textbf{\makecell{\acs{QRC}\\reviews}}} &
\cite{MANBGEtAl:21:AQT}: \acs{QRC} and \acs{QELM} formulations, input and output
settings, quantum substrates, measurement, and tasks &
$\checkmark$ & $\checkmark$ & $\checkmark$ & $\circ$ & $\checkmark$ & $\times$ & $\circ$ \\
&
\cite{FN:20:arxiv}: Pedagogical chapter on \acs{QRC}, \acs{QELM}, and quantum
circuit learning for near term devices &
$\checkmark$ & $\circ$ & $\circ$ & $\circ$ & $\circ$ & $\times$ & $\times$ \\
&
\cite{GWPSSEtAl:25:arxiv}: Quantum feature maps, a neutral atom \acs{QRC}
workflow, measurement, and applications &
$\circ$ & $\circ$ & $\checkmark$ & $\circ$ & $\circ$ & $\circ$ & $\circ$ \\
\midrule
\multirow{3}{0.105\textwidth}{\centering\textbf{\makecell{Physical\\\acs{RC}}}} &
\cite{AGM:24:Dy}: Physical reservoirs and selected quantum measurement
protocols for onboard artificial intelligence &
$\circ$ & $\times$ & $\circ$ & $\checkmark$ & $\checkmark$ & $\times$ & $\circ$ \\
&
\cite{TYRNKEtAl:19:NN}: Physical \acs{RC} organized by substrate, hardware,
applications, and practical constraints &
$\checkmark$ & $\times$ & $\circ$ & $\checkmark$ & $\checkmark$ & $\times$ & $\circ$ \\
&
\cite{ASAPSEtAl:25:JPP}: Photonic \acs{RC} hardware, input and output layers,
learning methods, and tasks &
$\circ$ & $\times$ & $\circ$ & $\checkmark$ & $\checkmark$ & $\times$ & $\circ$ \\
\midrule
\multirow{3}{0.105\textwidth}{\centering\textbf{\makecell{General\\\acs{RC}}}} &
\cite{YHBTLEtAl:24:NC}: \acs{RC} theory, algorithms, physical implementations,
applications, and adoption &
$\checkmark$ & $\times$ & $\circ$ & $\checkmark$ & $\checkmark$ & $\circ$ & $\circ$ \\
&
\cite{WTS:25:IJPEDS}: Benchmark taxonomy, task design, evaluation practice, and
limitations &
$\circ$ & $\times$ & $\circ$ & $\circ$ & $\checkmark$ & $\times$ & $\checkmark$ \\
&
\cite{ZV:23:IEEE_O_ACC}: \acs{RC} architectures, physical implementations,
software tools, and interdisciplinary applications &
$\checkmark$ & $\times$ & $\circ$ & $\checkmark$ & $\checkmark$ & $\circ$ & $\circ$ \\
\midrule
\centering\textbf{This survey} &
\acs{QRC} system model and taxonomy linking protocols, hardware, applications,
software and \acs{HPC}, and benchmarking &
$\checkmark$ & $\checkmark$ & $\checkmark$ & $\checkmark$ & $\checkmark$ & $\checkmark$ & $\checkmark$ \\
\bottomrule
\end{tabularx}
\smallskip

\noindent\footnotesize
$\checkmark$ = Sustained or dedicated coverage \quad
$\circ$ = Partial or contextual coverage \quad
$\times$ = No substantive coverage. Symbols compare topical coverage, not
quality. Fnd. = \acs{RC} or \acs{QRC} foundations; Tax. = \acs{QRC} taxonomy;
P/R = encoding, measurement, and readout protocols; HW = physical hardware;
SW = software or \acs{HPC} workflows; Eval. = benchmarking, resource
accounting, or reproducibility.
\end{table*}

\subsection{Contributions and Organization}
\label{sec:contributions}

This survey makes four contributions that correspond to the main analytical
stages of the paper.

\begin{itemize}[leftmargin=*]
\item \textbf{Preliminaries:} A common notation and system model that connects input preprocessing,
quantum encoding, reservoir dynamics, measurement, observable extraction, and
classical readout.
\item \textbf{\acs{QRC} Design:} A taxonomy and comparative analysis that connect
memory, nonlinearity, expressivity, and stability to architecture choices,
including temporal state location, encoding, measurement, feedback, and
trainable components.
\item \textbf{Implementation:} A synthesis of physical platforms and application
studies that distinguishes theoretical analysis, simulation, and hardware
experiments.
\item \textbf{Benchmarking:} A framework that connects software and \ac{HPC}
workflows, computational scaling, resource accounting, matched classical
baselines, and reproducibility requirements.
\end{itemize}

\begin{figure*}[t]
  \centering
  \includegraphics[width=\textwidth]{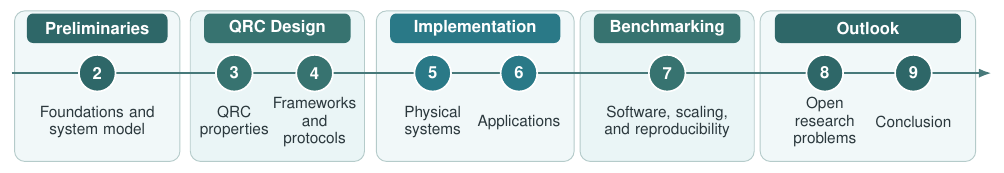}
  \caption{Organization and analytical flow of the survey.}
  \Description{Horizontal milestone diagram with five stages. Preliminaries
  contains Section 2. QRC Design contains Sections 3 and 4. Implementation
  contains Sections 5 and 6. Benchmarking contains Section 7. Outlook contains
  Sections 8 and 9. A continuous arrow connects the stages from left to right.}
  \label{fig:paper-outline}
\end{figure*}

As shown in Figure~\ref{fig:paper-outline}, the survey is organized into five
stages. Section~2 establishes the preliminaries and common system model.
Sections~3 and~4 examine \ac{QRC} design and operating protocols. Sections~5
and~6 cover physical implementations and applications. Section~7 addresses
software, computational scaling, resource accounting, and reproducibility.
Sections~8 and~9 present the open research problems and conclusions.

\section{Foundations and System Model}
\label{sec:foundations_system_model}

\begin{figure*}[t]
  \centering
  \includegraphics[width=\textwidth]{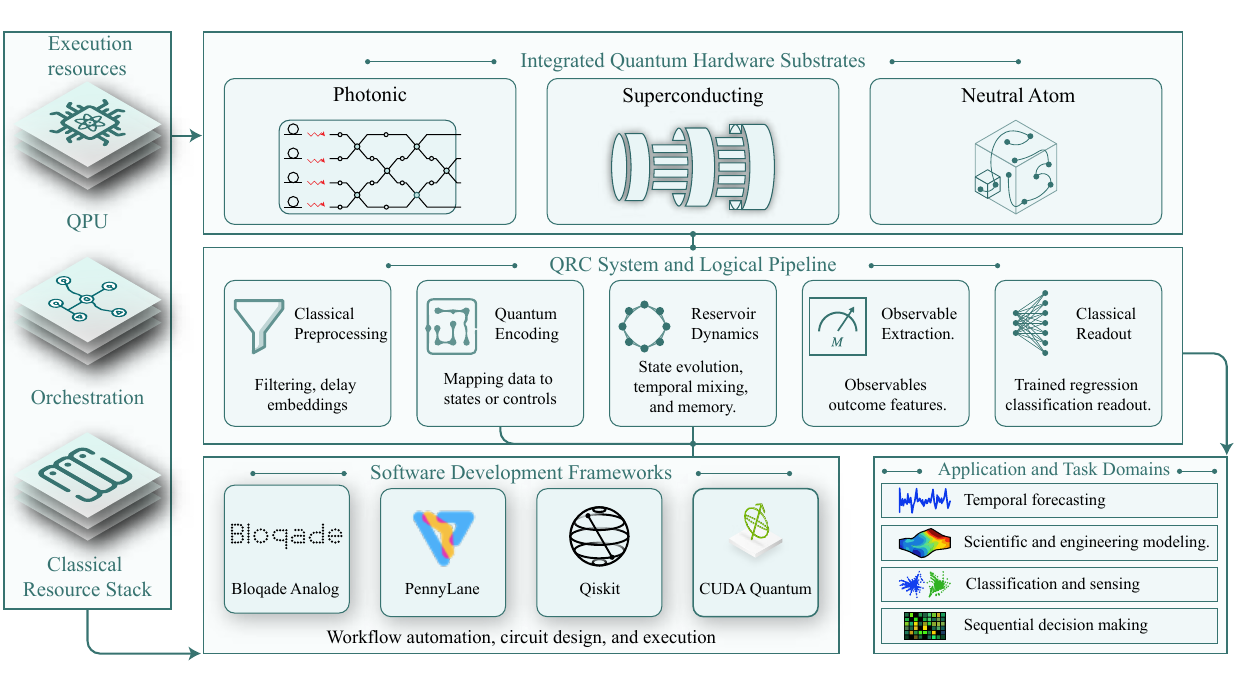}
  \caption{Integrated \ac{QRC} stack. Classical preprocessing maps
  application data to inputs accepted by the quantum interface. Reservoir
  evolution generates a state trajectory, measurement converts selected
  properties of that trajectory into classical features, and the readout maps
  those features to predictions. Finite sampling and measurement disturbance
  depend on the operating protocol. The surrounding blocks show
  representative hardware, software, execution resources, and task domains.}
  \Description{An integrated stack diagram for quantum reservoir computing.
  Classical preprocessing and quantum encoding feed a reservoir dynamics
  block, followed by observable extraction and a classical readout. The
  diagram also shows representative photonic, superconducting, and neutral
  atom hardware, software frameworks, hybrid execution resources, and
  application domains.}
  \label{fig:qrc-integrated-stack}
\end{figure*}

Interpreting a \ac{QRC} result requires an explicit account of every
transformation between the input and the prediction. We therefore define a
common system model that follows an ordered classical input through
preprocessing, quantum encoding, reservoir evolution, measurement, finite
sampling, and a classical readout. For a reader familiar with machine
learning, the reservoir state is analogous to a hidden state and the measured
quantities form a feature vector. The analogy is operational: the quantum
state cannot be read directly, and each accessible feature is determined by a
measurement protocol. In Figure~\ref{fig:qrc-integrated-stack}, we show how
these processing stages connect the hardware, software, computing, and
application layers. We illustrate the model with a spin based reservoir when
defining the encoding, evolution, measurement, and readout stages
\cite{FN:17:PRA_2}.

\subsection{QRC Model}
\label{subsec:qrc_system_model}

A data sequence of length $T$ is written as
\begin{align}
  \pmb{X}
  &=
  \left[
    \pmb{x}_{1},\ldots,\pmb{x}_{T}
  \right]
  \in \mathbb{R}^{D\times T},
  &
  \pmb{x}_{t}
  &\in \mathbb{R}^{D},
  \label{eq:qrc_input_sequence}
\end{align}
where $t\in\{1,\ldots,T\}$ is the input index and $D$ is the number of real
valued variables presented at each index. Let $\pmb{\lambda}_{p}$ contain the
preprocessing parameters, and let $\widetilde{D}$ be the processed input
dimension. Causal preprocessing uses only the current and earlier inputs:
\begin{align}
  \widetilde{\pmb{x}}_{t}
  =
  \mc{P}_{\pmb{\lambda}_{p}}
  \left(
    \pmb{x}_{1},\ldots,\pmb{x}_{t}
  \right)
  \in
  \mathbb{R}^{\widetilde{D}}.
  \label{eq:qrc_preprocessing_map}
\end{align}
If $\mc{P}$ uses a delay buffer, running normalization, or another classical
state, that state belongs to the preprocessing layer and is distinct from the
retained quantum state.

Let $\mc{H}_{R}$ denote the reservoir Hilbert space. A density operator
$\pmb{\rho}$ on $\mc{H}_{R}$ is Hermitian, positive semidefinite, and has unit
trace. We write the set of such operators as
$\mc{D}(\mc{H}_{R})$. These conditions ensure real nonnegative measurement
probabilities whose total is one. The reservoir begins in
$\pmb{\rho}_{0}\in\mc{D}(\mc{H}_{R})$. For $N$ qubits,
\begin{align}
  \mc{H}_{R}
  &=
  \left(\mathbb{C}^{2}\right)^{\otimes N},
  &
  d_{R}
  &=
  \dim(\mc{H}_{R})
  =
  2^{N}.
  \label{eq:qrc_qubit_hilbert_space}
\end{align}
Qudit, \ac{CV}, and hybrid reservoirs use different state spaces and operator
algebras. In every case, $d_R$ describes the state space dimension rather than
the number of features available to the readout
\cite{NAGPSEtAl:21:CP,SPKRCEtAl:24:NC}.

Within one input interval, the reservoir can be sampled at $V\geq1$
intermediate times called virtual nodes. The index $v=0$ denotes the state
immediately after encoding, and $v\in\{1,\ldots,V\}$ denotes the state after
the corresponding evolution interval. Thus, $\pmb{\rho}_{t,v}$ is the state
associated with input $t$ and virtual node $v$, while
$\pmb{\rho}_{t,V}$ is the state retained for the next input. Measurement
produces a feature vector of fixed dimension
$\pmb{z}_{t}\in\mathbb{R}^{F}$, and the classical readout returns a prediction
$\widehat{\pmb{y}}_{t}\in\mathbb{R}^{D_y}$, where $F$ is the feature count and
$D_y$ is the output dimension.

\begin{definition}[QRC predictor]
\label{def:qrc_model}
For a processed input $\widetilde{\pmb{x}}_t$ and a retained state
$\pmb{\rho}_{t-1,V}$, a \ac{QRC} predictor is the possibly stochastic
algorithmic map
\begin{align}
  \mc{Q}:
  \left(
    \widetilde{\pmb{x}}_{t},
    \pmb{\rho}_{t-1,V}
  \right)
  \longmapsto
  \left(
    \left\{\pmb{\rho}_{t,v}\right\}_{v=0}^{V},
    \pmb{z}_{t},
    \widehat{\pmb{y}}_{t}
  \right).
  \label{eq:qrc_formal_definition}
\end{align}
The map composes the encoding, reservoir evolution, measurement, and
classical readout defined below. It is not a quantum channel because its
outputs include classical data. It can be stochastic because finite
measurement samples and conditional outcomes are random.
\end{definition}

In the conventional setting, the reservoir configuration is fixed while the
readout is trained. Some variants select encoding, dynamics, or measurement
parameters in an outer model selection loop
\cite{FN:17:PRA_2,CNY:20:PRA_2,MANBGEtAl:21:AQT}. If no quantum state is
retained between inputs, $\pmb{\rho}_{t-1,V}$ in
Definition~\ref{def:qrc_model} is replaced by a fixed reference state. The
quantum component then acts as a memoryless feature map, as in a \ac{QELM}.

For example, a spin based reservoir \cite{FN:17:PRA_2} replaces one qubit
state with a scalar input, while the remaining qubits retain temporal
information. A fixed interacting spin Hamiltonian evolves the full
state, single qubit observables are sampled at several virtual nodes, and a
linear readout maps the resulting real vector to the target.

\subsection{Preprocessing and Quantum Encoding}
\label{subsec:qrc_preprocessing_encoding}

The preprocessing map in Eq.~\eqref{eq:qrc_preprocessing_map} adapts the data
to the controls available at the quantum interface. Its parameters can specify
scaling, filtering, feature selection, or dimension reduction. These
parameters must be fitted on training data and then held fixed for validation
and testing. Range normalization is widely used because
rotation angles, detunings, drives, displacements, and optical controls have
calibrated operating intervals. Image and field data may additionally
require downsampling, principal component projection, or modal reduction to
match the available controls
\cite{AHZWWEtAl:24:arxiv,TTCS:25:arxiv,PHS:23:PRR}.

For a temporal task, preprocessing can construct the explicit delay window
\begin{align}
  \pmb{w}_{t}
  =
  \left[
    \pmb{x}_{t-L+1}^{\mathrm{T}},
    \ldots,
    \pmb{x}_{t}^{\mathrm{T}}
  \right]^{\mathrm{T}},
  \label{eq:qrc_preprocessing_window}
\end{align}
where $L$ is the number of included input indices. Values preceding the
recorded sequence require a declared padding or initialization rule. In a
recurrent reservoir, the current sample can be encoded while the retained
quantum state carries history. In recurrence free protocols, the delay window
supplies temporal context to the predictor, not to the quantum state itself
\cite{AHZWWEtAl:24:arxiv,MGOEZEtAl:25:CIJNS,MHZBH:26:arxiv}.

A quantum channel is a linear map on density operators that is trace
preserving and completely positive. Trace preservation keeps total
probability equal to one. Complete positivity ensures that the map remains
valid when the reservoir is part of a larger quantum system. Let
$\pmb{\lambda}_{e}$ contain all fixed or selected encoding parameters. The
encoding channel is
\begin{align}
  \mc{E}_{\widetilde{\pmb{x}};\pmb{\lambda}_{e}}
  &:
  \mc{D}(\mc{H}_{R})
  \rightarrow
  \mc{D}(\mc{H}_{R}),
  \label{eq:qrc_encoding_map_general}
  \\
  \pmb{\rho}_{t,0}
  &=
  \mc{E}_{\widetilde{\pmb{x}}_{t};\pmb{\lambda}_{e}}
  \left(
    \pmb{\rho}_{t-1,V}
  \right),
  \qquad
  \pmb{\rho}_{0,V}=\pmb{\rho}_{0}.
  \label{eq:qrc_input_injection}
\end{align}
Equation~\eqref{eq:qrc_input_injection} also defines
$\pmb{\rho}_{t,0}$ as the state at virtual node zero.

One recurrent encoding preserves part of the previous state. Split the
reservoir into an input subsystem $\mc{H}_{\mathrm{in}}$ and a retained
subsystem $\mc{H}_{\mathrm{mem}}$, so that
$\mc{H}_{R}=\mc{H}_{\mathrm{in}}\otimes\mc{H}_{\mathrm{mem}}$. Let
$\pmb{\rho}_{\mathrm{in}}(\widetilde{\pmb{x}}_t)$ be the state prepared from
the input, and let $\Tr_{\mathrm{in}}$ denote the partial trace over the old
input subsystem. Subsystem replacement acts as
\begin{align}
  \mc{E}_{\widetilde{\pmb{x}}_{t};\pmb{\lambda}_{e}}
  \left(\pmb{\rho}\right)
  =
  \pmb{\rho}_{\mathrm{in}}
  \left(\widetilde{\pmb{x}}_{t}\right)
  \otimes
  \Tr_{\mathrm{in}}\!\left[\pmb{\rho}\right].
  \label{eq:qrc_subsystem_replacement}
\end{align}
The input dependent density operator replaces $\mc{H}_{\mathrm{in}}$, while
the reduced state on $\mc{H}_{\mathrm{mem}}$ is preserved.

For example, a spin based reservoir \cite{FN:17:PRA_2} can encode a scalar
$s_t\in[0,1]$ as
\begin{align}
  \pmb{\rho}_{\mathrm{in}}(s_t)
  &=
  \ket{\psi(s_t)}\!\bra{\psi(s_t)},
  &
  \ket{\psi(s_t)}
  &=
  \sqrt{1-s_t}\ket{0}+\sqrt{s_t}\ket{1}.
  \label{eq:qrc_spin_encoding_example}
\end{align}
This is state preparation by subsystem replacement, not amplitude encoding of
a classical vector. Circuits instead use input dependent
gates, often with reset and repeated data upload on selected qubits
\cite{YSKNGEtAl:23:arxiv,HCLAW:25:arxiv,ATM:25:QMI}.

\begin{table*}[t]
\centering
\caption{Input and control interfaces used in \ac{QRC} implementations.
The listed controls and settings are representative examples.}
\label{tab:qrc_encoding_modalities}
\footnotesize
\setlength{\tabcolsep}{3.2pt}
\renewcommand{\arraystretch}{1.16}
\begin{tabularx}{\textwidth}{@{}p{0.22\textwidth}p{0.22\textwidth}p{0.27\textwidth}X@{}}
\toprule
\rowcolor{black!5}
\textbf{Interface} &
\makecell[l]{\textbf{Encoded}\\\textbf{quantity}} &
\makecell[l]{\textbf{Representative}\\\textbf{controls}} &
\makecell[l]{\textbf{Representative}\\\textbf{settings}} \\
\midrule
Subsystem state replacement
\cite{FN:17:PRA_2,AKS:25:arxiv,MHZBH:26:arxiv} &
Input density operator &
Prepared input subsystem and replacement schedule &
Spin networks and recurrent many body models \\
\addlinespace
Circuit gates
\cite{YSKNGEtAl:23:arxiv,HCLAW:25:arxiv,ATM:25:QMI} &
Encoding unitary &
Rotation angles, data upload, and input qubit reset &
Digital and superconducting circuits \\
\addlinespace
Analog Hamiltonian control
\cite{AHZWWEtAl:24:arxiv,DAW:25:arxiv,VVVTMEtAl:25:arxiv} &
Control waveform &
Detuning, Rabi drive, geometry, and interaction scale &
Neutral atom and related analog devices \\
\addlinespace
Bosonic and \acs{CV} control
\cite{NAGPSEtAl:21:CP,CDBGM:26:arxiv,SPKRCEtAl:24:NC} &
Mode state or drive &
Displacement, squeezing, quadrature drive, and phase &
Oscillators, cavities, and Gaussian networks \\
\addlinespace
Photonic multiplexing
\cite{ALFDAEtAl:25:NC,CSPMMEtAl:26:NQI} &
Optical pulse or mode transformation &
Amplitude, wavelength, delay, and optical phase &
Delay loops and multimode optical systems \\
\bottomrule
\end{tabularx}
\end{table*}

\begin{samepage}
Analog implementations can encode data through controlled dynamics. Let
$\pmb{\lambda}_{H}$ contain the fixed or selected dynamics parameters, let
$\pmb{H}_{0}$ be the drift Hamiltonian, and let
$\{\pmb{H}_{q}\}_{q=1}^{Q}$ be the $Q$ available control Hamiltonians. If
$\tau$ denotes continuous time within the input interval and $\phi_q$ maps the
processed input to control waveform $q$, then
\begin{align}
  \pmb{H}
  \left(
    \widetilde{\pmb{x}}_{t},\tau
  \right)
  =
  \pmb{H}_{0}
  \left(
    \tau;\pmb{\lambda}_{H}
  \right)
  +
  \sum_{q=1}^{Q}
  \phi_{q}
  \left(
    \widetilde{\pmb{x}}_{t},\tau;
    \pmb{\lambda}_{e}
  \right)
  \pmb{H}_{q},
  \label{eq:qrc_encoding_class_hamiltonian}
\end{align}
Here, $\pmb{\lambda}_{e}$ also includes parameters of the input dependent
waveforms. The available controls depend on the substrate. For example,
neutral atom positions normally specify a device geometry, whereas detuning
and Rabi waveforms can change between inputs.
Table~\ref{tab:qrc_encoding_modalities} lists representative interfaces. Its
rows are examples rather than mutually exclusive categories.
\end{samepage}

\subsection{Reservoir Evolution}
\label{subsec:qrc_evolution}

The reservoir evolves the encoded state between virtual nodes. For each
$v\in\{1,\ldots,V\}$, the channel
\begin{align}
  \mc{R}_{v;\pmb{\lambda}_{H}}
  &:
  \mc{D}(\mc{H}_{R})
  \rightarrow
  \mc{D}(\mc{H}_{R}),
  \label{eq:qrc_reservoir_map_general}
  \\
  \pmb{\rho}_{t,v}
  &=
  \mc{R}_{v;\pmb{\lambda}_{H}}
  \left(
    \pmb{\rho}_{t,v-1}
  \right).
  \label{eq:qrc_state_evolution_general}
\end{align}
maps the previous density operator to the next one. The index $v$ identifies
the sampling interval, and $\pmb{\lambda}_{H}$ contains the dynamics
parameters defined above. This \ac{CPTP} description covers closed evolution,
noisy circuits, and Markovian open evolution integrated over a finite
interval. It is nonselective because no measurement outcome is retained in
the state update. Conditional evolution is defined in
Section~\ref{subsec:qrc_measurement}.

The channel that propagates the state from virtual node zero to node $v$ is
\begin{align}
  \mc{R}_{v:1}
  =
  \mc{R}_{v;\pmb{\lambda}_{H}}
  \circ\cdots\circ
  \mc{R}_{1;\pmb{\lambda}_{H}}.
  \label{eq:qrc_cumulative_channel}
\end{align}
where composition is evaluated from right to left. If the Hamiltonian
$\pmb{H}_{v}$ is constant during interval $v$, closed evolution takes the
unitary form
\begin{align}
  \pmb{\rho}_{t,v}
  &=
  \pmb{U}_{v}
  \pmb{\rho}_{t,v-1}
  \pmb{U}_{v}^{\dagger},
  &
  \pmb{U}_{v}
  &=
  \exp
  \left(
    -\mathrm{i}\pmb{H}_{v}\Delta\tau_{v}
  \right),
  \label{eq:qrc_virtual_node_state_recursive}
\end{align}
where $\Delta\tau_v$ is the duration of interval $v$, $\mathrm{i}$ is the
imaginary unit, and units with $\hbar=1$ are used. Let $\tau_v$ be the end of
interval $v$, so $\Delta\tau_v=\tau_v-\tau_{v-1}$. A time dependent
Hamiltonian instead gives
\begin{align}
  \pmb{U}_{v}
  =
  \mc{T}
  \exp
  \left[
    -\mathrm{i}
    \int_{\tau_{v-1}}^{\tau_v}
    \pmb{H}
    \left(
      \widetilde{\pmb{x}}_{t},\tau
    \right)
    d\tau
  \right],
  \label{eq:qrc_time_ordered_unitary}
\end{align}
where $\mc{T}$ orders operators according to time.

For example, a spin based reservoir \cite{FN:17:PRA_2} can use the fixed
Hamiltonian
\begin{align}
  \pmb{H}_{\mathrm{spin}}
  =
  \sum_{i<j}J_{ij}\pmb{X}_{i}\pmb{X}_{j}
  +
  h\sum_{i=1}^{N}\pmb{Z}_{i},
  \label{eq:qrc_spin_hamiltonian_example}
\end{align}
where $\pmb{X}_{i}$ and $\pmb{Z}_{i}$ are Pauli operators acting on qubit
$i$, $J_{ij}$ are real spin couplings, and $h$ is the transverse field. The
input interval of duration $\tau$ is divided into $V$ equal intervals, so
$\Delta\tau_v=\tau/V$. The same unitary law is used at each interval, while
sampling at the intermediate times creates the virtual nodes.

Markovian open reservoirs combine coherent evolution with loss, relaxation,
or pumping. Under a time local completely positive semigroup model, their
generator has the Gorini Kossakowski Sudarshan Lindblad form
\begin{align}
  \frac{d\pmb{\rho}}{d\tau}
  =
  -\mathrm{i}
  \left[
    \pmb{H},\pmb{\rho}
  \right]
  +
  \sum_{k}
  \gamma_{k}
  \left(
    \pmb{L}_{k}\pmb{\rho}\pmb{L}_{k}^{\dagger}
    -
    \frac{1}{2}
    \left\{
      \pmb{L}_{k}^{\dagger}\pmb{L}_{k},
      \pmb{\rho}
    \right\}
  \right),
  \label{eq:lindblad}
\end{align}
where $k$ indexes dissipative processes, $\pmb{L}_{k}$ is a jump operator,
$\gamma_k\geq0$ is its rate, and
$\{A,B\}=AB+BA$ denotes the anticommutator. The Markovian assumption excludes
general environment memory. Integrating Eq.~\eqref{eq:lindblad} over one
sampling interval gives a reservoir channel. Dissipation can contract
differences between initial states and thereby support the echo state
property. Its rate must be considered together with the input interval and
task. Rapid contraction can remove useful dependence on recent inputs, while
weak contraction can preserve unwanted dependence on the initialization
\cite{CN:19:QIP_2,SASGZ:24:Qu,AO:25:PRE}.

The final state $\pmb{\rho}_{t,V}$ closes the recurrence by becoming the input
state at time $t+1$. This quantum state is not the only possible location of
temporal context. Other protocols use a measurement record, a feedback loop,
an optical delay, or an explicit classical window. These mechanisms use
different resources and are distinguished in
Section~\ref{sec:qrc_framework_categories}
\cite{FN:17:PRA_2,CN:19:QIP_2,MAGSZ:23:NQI,AHZWWEtAl:24:arxiv}.

Spin networks, neutral atom arrays, bosonic modes, fermionic hopping models,
photonic systems, and hybrid light matter devices instantiate these maps on
different state spaces and operator algebras. These descriptions overlap:
a neutral atom array is also a spin system, while chaos is a dynamical regime
rather than a substrate. Particle statistics belong to the substrate
specification, not to the tunable vector $\pmb{\lambda}_{H}$. Platform
Hamiltonians, operating assumptions, and hardware constraints are examined
in Section~\ref{sec:qrc_hardware}
\cite{LCGZ:23:AQT,BNGY:22:PQ,NAGPSEtAl:21:CP,DGZ:26:PRR}.

\subsection{Measurement, Sampling, and Conditional Feedback}
\label{subsec:qrc_measurement}

The measurement layer determines which properties of the state trajectory are
available to the classical readout. Let $M$ be the number of Hermitian
observables and let $\pmb{\lambda}_{M}$ collect the measurement choices, such
as the observables, sampled nodes, and shot allocation. The observable set is
\begin{align}
  \mc{O}_{\pmb{\lambda}_{M}}
  =
  \left\{
    \pmb{O}_{m}
  \right\}_{m=1}^{M}.
  \label{eq:qrc_observable_set}
\end{align}
At virtual node $v$, the exact expectation of observable $m$ is
\begin{align}
  z_{t,m,v}
  =
  \Tr
  \left[
    \pmb{O}_{m}\pmb{\rho}_{t,v}
  \right],
  \qquad
  m=1,\ldots,M,
  \quad
  v=1,\ldots,V.
  \label{eq:qrc_observable_expectation}
\end{align}
The trace $\Tr[\pmb{O}_{m}\pmb{\rho}_{t,v}]$ is real because
$\pmb{O}_{m}$ and $\pmb{\rho}_{t,v}$ are Hermitian. Concatenating the values
for all $M$ observables and $V$ nodes gives the ideal vector
$\pmb{z}^{\mathrm{obs}}_{t}\in\mathbb{R}^{MV}$.

\begin{samepage}
\begin{definition}[Ideal and sampled reservoir features]
\label{def:qrc_features}
An ideal reservoir feature is an exact expectation or outcome probability
computed from a density operator. A sampled reservoir feature is an estimator
formed from a finite number of measurement outcomes. The ideal quantity
defines the target of estimation, whereas the sampled quantity is the value
available to a hardware readout.
\end{definition}
\end{samepage}

For example, a spin based reservoir \cite{FN:17:PRA_2} can use
$\pmb{O}_{i}=\pmb{Z}_{i}$ for $i\in\{1,\ldots,N\}$. Hence, $M=N$ and
concatenating all virtual nodes gives $F=NV$ ideal features before any
selection or conditioning. More generally, if $N_M\leq N$ qubits are measured, the
single qubit Pauli operators $\pmb{\sigma}_{i}^{\alpha}$ with
$\alpha\in\{x,y,z\}$ provide at most $3N_M$ candidates. Products of Pauli
operators expose correlations. Their execution cost depends on which
operators can share a measurement setting and on the required precision.
Random measurement constructions provide other feature families, while
concentration can make observable values statistically difficult to
distinguish as the system grows
\cite{TFHS:25:MLST_2,XHASCEtAl:25:arxiv,SGZ:25:arxiv}.

Hardware returns outcomes rather than exact expectations. Suppose
$\pmb{\sigma}_{i}^{\alpha}$ is measured independently on $S$ copies of the
same state at node $(t,v)$. If $s\in\{1,\ldots,S\}$ indexes the copy and
$\lambda^{(s)}_{t,i,\alpha,v}\in\{-1,+1\}$ is its outcome, write the exact
expectation as $z_{t,i,\alpha,v}$. The estimator and its variance are
\begin{align}
  \widehat{z}_{t,i,\alpha,v}
  &=
  \frac{1}{S}
  \sum_{s=1}^{S}
  \lambda^{(s)}_{t,i,\alpha,v},
  &
  \operatorname{Var}
  \left(
    \widehat{z}_{t,i,\alpha,v}
  \right)
  &=
  \frac{
    1-z_{t,i,\alpha,v}^{2}
  }{S},
  \label{eq:pauli_shot_estimator}
\end{align}
where $\widehat{z}_{t,i,\alpha,v}$ is the sampled estimate. A larger $S$
reduces variance but increases quantum executions. If measurement destroys a
recurrent state, the copies must be synchronized before measurement or the
relevant input prefix must be replayed. Reusing the measured trajectory
instead changes the subsequent state dynamics. Reset removes intrinsic
quantum state memory unless another component supplies temporal context
\cite{ATM:25:QMI,MAGSZ:23:NQI,FOLNM:25:PRX10}.

An observable expectation is not the only possible feature. A positive
operator valued measure, or POVM, assigns probabilities to a finite or
countable outcome set $\mc{Y}$. Its effects
$\{\pmb{E}_{r}\}_{r\in\mc{Y}}$ satisfy
$\pmb{E}_{r}\succeq0$ and
$\sum_{r\in\mc{Y}}\pmb{E}_{r}=\pmb{I}_{R}$, where $\pmb{I}_{R}$ is the
identity operator on $\mc{H}_{R}$. The probability of outcome $r$ at node
$(t,v)$ is
\begin{align}
  p_{t,r,v}
  =
  \Tr
  \left[
    \pmb{E}_{r}\pmb{\rho}_{t,v}
  \right].
  \label{eq:qrc_povm_probability}
\end{align}
A readout can use this probability vector. For a declared real valued function
$g_m$ of the outcome, it can instead use the fixed statistic
$\sum_{r\in\mc{Y}}g_m(r)p_{t,r,v}$. A POVM specifies probabilities but does not specify
the state after measurement. A quantum instrument is required when
measurement disturbance or outcome dependent feedback affects later inputs.

To describe a measurement after the final virtual node, let
$\{\mc{M}^{(t)}_{r}\}_{r\in\mc{Y}}$ be completely positive operations whose
sum is trace preserving. This collection is a quantum instrument. The state
before measurement and the unnormalized state associated with outcome $r$ are
\begin{align}
  \pmb{\rho}^{-}_{t,V}
  &=
  \left(
    \mc{R}_{V:1}
    \circ
    \mc{E}_{\widetilde{\pmb{x}}_{t};\pmb{\lambda}_{e}}
  \right)
  \left(
    \pmb{\rho}_{t-1,V}
  \right),
  \label{eq:qrc_premeasurement_state}
  \\
  \widetilde{\pmb{\rho}}_{t,r}
  &=
  \mc{M}^{(t)}_{r}
  \left(
    \pmb{\rho}^{-}_{t,V}
  \right).
  \label{eq:qrc_instrument_state}
\end{align}
Let $\mc{M}^{(t)\dagger}_{r}$ denote the adjoint operation, defined by
$\Tr[\pmb{A}\mc{M}^{(t)}_{r}(\pmb{B})]
=\Tr[\mc{M}^{(t)\dagger}_{r}(\pmb{A})\pmb{B}]$ for operators $\pmb{A}$ and
$\pmb{B}$. The instrument induces the POVM effect and probability
\begin{align}
  \pmb{E}^{(t)}_{r}
  &=
  \mc{M}^{(t)\dagger}_{r}
  \left(
    \pmb{I}_{R}
  \right),
  &
  p_t(r)
  &=
  \Tr
  \left[
    \widetilde{\pmb{\rho}}_{t,r}
  \right]
  =
  \Tr
  \left[
    \pmb{E}^{(t)}_{r}\pmb{\rho}^{-}_{t,V}
  \right].
  \label{eq:qrc_instrument_born}
\end{align}
If outcome $r_t$ occurs and $p_t(r_t)>0$, its normalized conditional state can
be passed through a reset or feedback channel $\mc{F}_{r_t}$:
\begin{align}
  \pmb{\rho}_{t,V}
  =
  \mc{F}_{r_t}
  \left(
    \frac{
      \widetilde{\pmb{\rho}}_{t,r_t}
    }{
      p_t(r_t)
    }
  \right),
  \label{eq:qrc_feedback_update}
\end{align}
Each $\mc{F}_{r}$ is a specified \ac{CPTP} map selected by outcome $r$.
Protocols that measure at every virtual node require an instrument indexed by
both $t$ and $v$ and an outcome record with both indices
\cite{MAGSZ:23:NQI,KFY:24:PQ,FOLNM:25:PRX10}.

A readout requires a fixed dimensional input unless it is itself recurrent.
It can therefore use the current outcome, a window
$r_{t-L_r+1:t}$ containing $L_r$ outcomes, or a statistic of declared
dimension $D_r$ formed by a function $\psi$:
\begin{align}
  \pmb{h}_{t}
  =
  \psi
  \left(
    r_{t-L_r+1:t}
  \right)
  \in
  \mathbb{R}^{D_r},
  \label{eq:qrc_instrument_features}
\end{align}
The window needs an initialization rule when $t<L_r$. Using the full growing
history requires an explicitly recurrent classical readout. If outcomes are
discarded, the nonselective channel is
$\sum_{r}\mc{M}^{(t)}_{r}$. With outcome dependent feedback it becomes
$\sum_{r}\mc{F}_{r}\circ\mc{M}^{(t)}_{r}$. Both maps can disturb the retained
state even when the outcome is not supplied to the readout.

Conditioning applied after quantum evaluation belongs to the measured feature
pipeline. One finite sampling study applies singular value decomposition,
truncation, and smoothing to activation matrices after measurement. Its
reported gains depend on the comparison: truncation can underperform the
full noisy matrix while outperforming a noisy matrix restricted to the same
rank \cite{ATM:25:QMI}. For a fixed reservoir and training set, a separate
study reports a supervised kernel ridge construction for selecting an
observable. Its decomposition or eigenbasis can guide implementation, but
measurement settings, shots, and optimization data remain part of the
resource account \cite{GR:26:arxiv}.

\subsection{Classical Readout and Model Selection}
\label{subsec:qrc_readout}

The readout receives the fixed dimensional vector
$\pmb{z}_{t}\in\mathbb{R}^{F}$ defined in
Definition~\ref{def:qrc_model}. In an ideal observable protocol,
$F=MV$ and $\pmb{z}_{t}=\pmb{z}^{\mathrm{obs}}_{t}$. A hardware protocol
replaces exact expectations by sampled estimates. A conditional protocol can
also append the statistic $\pmb{h}_{t}$. Feature scaling, selection, and
conditioning must be fitted using only the training data and then fixed for
validation and testing.

Let $f_{\pmb{\theta}_{l}}$ be the classical readout with trainable parameters
$\pmb{\theta}_{l}$. Given a target
$\pmb{y}_{t}\in\mathbb{R}^{D_y}$, its prediction is
\begin{align}
  \widehat{\pmb{y}}_{t}
  =
  f_{\pmb{\theta}_{l}}
  \left(
    \pmb{z}_{t}
  \right).
  \label{eq:qrc_classical_readout}
\end{align}
The standard choice is a linear readout. Adding a leading one to the feature
vector incorporates an intercept:
\begin{align}
  \widehat{\pmb{y}}_{t}
  &=
  \pmb{W}_{\mathrm{out}}
  \widetilde{\pmb{z}}_{t},
  &
  \widetilde{\pmb{z}}_{t}
  &=
  \left[
    1,
    \pmb{z}_{t}^{\mathrm{T}}
  \right]^{\mathrm{T}},
  &
  \pmb{W}_{\mathrm{out}}
  &\in
  \mathbb{R}^{D_y\times(F+1)}.
  \label{eq:qrc_linear_readout}
\end{align}
Here, $\widetilde{\pmb{z}}_{t}\in\mathbb{R}^{F+1}$ is the augmented feature
vector and $\pmb{W}_{\mathrm{out}}$ contains the trainable weights.

A recurrent reservoir is commonly run for a declared washout interval before
training samples are collected. After that interval, let
$\pmb{Z}_{\mathrm{tr}}\in\mathbb{R}^{(F+1)\times K_{\mathrm{tr}}}$ contain
$K_{\mathrm{tr}}$ training feature columns and let
$\pmb{Y}_{\mathrm{tr}}\in\mathbb{R}^{D_y\times K_{\mathrm{tr}}}$ contain
their targets. With $\pmb{I}$ denoting the identity matrix of size $F+1$,
ridge regression with penalty $\beta>0$ gives
\begin{align}
  \pmb{W}_{\mathrm{out}}
  =
  \pmb{Y}_{\mathrm{tr}}
  \pmb{Z}_{\mathrm{tr}}^{\mathrm{T}}
  \left(
    \pmb{Z}_{\mathrm{tr}}
    \pmb{Z}_{\mathrm{tr}}^{\mathrm{T}}
    +
    \beta\pmb{I}
  \right)^{-1},
  \qquad
  \beta>0.
  \label{eq:qrc_readout_training}
\end{align}
For example, a spin based reservoir \cite{FN:17:PRA_2} can form the columns of
$\pmb{Z}_{\mathrm{tr}}$ from sampled or ideal $\pmb{Z}_{i}$ expectations
across all $N$ qubits and $V$ virtual nodes. Only $\pmb{W}_{\mathrm{out}}$ is
fitted in the conventional configuration.

Preprocessing, quantum interface, dynamics, measurement, and readout choices
form an outer model selection problem. Candidate parameters are fitted on the
training split, selected on a validation split, and evaluated once on a held
out test split.
Task metrics may be considered together with measurement precision, latency,
and total execution count
\cite{KFS:20:SR,GR:26:arxiv}. Resource accounting should include all quantum
runs used for parameter selection, measurement settings, shots, sequence
replay, and final testing, rather than only the executions used by the
reported readout.

The model separates the state space, input interface, evolution, measurement,
and readout while preserving their resource dependencies.
It supports direct comparison of recurrent and recurrence free protocols,
conditional and nonselective measurements, and distinct physical substrates
without treating these choices as interchangeable.
\section{Computational Properties and Design Tradeoffs of \ac{QRC}}
\label{sec:qrc_properties_design}
The computational value of a \ac{QRC} system depends on which distinctions
among input histories remain accessible after quantum evolution and
measurement. We characterize the predictor in
Definition~\ref{def:qrc_model} as a causal input and output filter formed by
preprocessing, quantum encoding, reservoir evolution, feature extraction, and
classical readout. For the ordered history
$\pmb{x}_{1:t}=(\pmb{x}_{1},\ldots,\pmb{x}_{t})$, this filter returns
$\widehat{\pmb{y}}_{t}$ from the ideal or sampled features distinguished in
Definition~\ref{def:qrc_features}. Its performance is governed by the composed
map and the accessible features, not by Hilbert space dimension alone. The
subsections examine echo state and fading memory properties; memory and
information processing capacity; nonlinearity and expressivity; accessible
quantum feature dimension; universality; and constraints imposed by noise,
finite sampling, and \ac{NISQ} hardware. A useful reservoir must suppress
dependence on its initial state while preserving measurable distinctions among
recent inputs.

\subsection{Echo state and fading memory properties}
The \emph{\ac{ESP}} requires the reservoir state to become independent of
initialization for each input sequence whose processed values lie in the
declared operating domain. In trace norm, a sufficient operational statement is
$\|\pmb{\rho}^{(a)}_{t,v}-\pmb{\rho}^{(b)}_{t,v}\|_{1}\to0$ as
$t\to\infty$ for any two initial density operators
$\pmb{\rho}^{(a)}_{0},\pmb{\rho}^{(b)}_{0}\in\mc{D}(\mc{H}_{R})$ driven by the same
inputs, with the virtual node index $v$ fixed. Here,
$\|A\|_{1}=\Tr\sqrt{A^{\dagger}A}$ is the trace norm. For two density operators,
one half of the trace norm of their difference is the trace distance and
quantifies their optimal distinguishability by measurement. The
\emph{\ac{FMP}} concerns the associated causal filter: it is continuous under a
weighted norm on past input sequences, so changes in the distant past have
progressively less influence on the output
\cite{GO:18:NN,KTN:24:PRE,CN:19:QIP_2}.
Contractive channels provide sufficient conditions for these properties in
specified quantum reservoir families, but contraction alone does not ensure
useful computation. Input histories relevant to the task must remain
distinguishable through the selected observables. Input dependence and
observability therefore complement forgetting conditions
\cite{AO:25:PRE,KTN:24:PRE}. Vector representations of quantum states connect
these recursions to systems with affine state updates and classical echo state theory
\cite{AO:23:PRE}. Dissipation, noise, measurement, and reset schedules can
change the contraction rate, but their useful range depends on the task and
protocol because rapid contraction can remove both initial state dependence
and desired recent memory \cite{MRNP:24:V_2IICQCEQ,MDP:23:PL}.

\subsection{Memory and information processing capacity}
Linear \ac{MC} measures how accurately a trained readout reconstructs a
delayed input at discrete lag $\ell$ from the current reservoir feature vector
$\pmb{z}_{t}$. \Ac{IPC} extends the target family to orthogonal nonlinear
functions of the input history \cite{DVSM:12:SR,LCGZ:23:AQT}. Both quantities
depend on the input distribution, target basis, measured feature set, readout
family, training length, regularization, and an estimator derived from finite
samples. They should therefore be
reported under a stated protocol rather than treated as intrinsic properties
of Hilbert space and reservoir dynamics alone. Capacity bounds refer to the
effective readout dimension under their assumptions, not to formal Hilbert
space dimension. Comparative studies show that
particle statistics, coupling topology, reset rate, and sampling interval can
redistribute linear and nonlinear capacities
\cite{LCGZ:23:AQT,MDP:23:PL,KFS:20:SR,IDJ:25:PRR,IJD:25:PRR}. Studies of
cavity quantum electrodynamics and discrete time crystal reservoirs provide
further examples in which
capacity changes with the operating regime
\cite{DGZ:26:PRR,ZLGYJEtAl:25:arxiv}. These comparisons show sensitivity to
design choices within the tested models, not a universal ranking of physical
substrates.

\subsection{Nonlinearity and expressivity}
A reservoir is useful when functions of the input history that are relevant to
the task are
linearly accessible to the readout. Quantum evolution is linear as a map on
density operators, but the complete map from classical data to measured
features can be nonlinear through state preparation, controls that depend on
the input, repeated injection, the dependence of measurement statistics on the
input, and fixed nonlinear postprocessing
\cite{GRRO:22:NCE,TTNLBEtAl:25:arxiv}. Reservoirs with the same
Hamiltonian can therefore have different expressivity when their encodings or
observable sets differ. Encoding, dynamics, measurement, and classical
postprocessing must be specified together. Nonlinear benchmark performance
does not establish a quantum advantage unless the comparison also controls
classical feature dimension, preprocessing, training, and measurement cost
\cite{TTNLBEtAl:25:arxiv,IJD:25:PRR}.

\subsection{Accessible quantum feature dimension}
Memory and information processing capacities assess reconstruction of
specified target families. Accessible feature dimension instead concerns the
number and conditioning of independent feature directions exposed by the
encoding, dynamics, and measurement. It depends on observable variability, the
rank and conditioning of the feature matrix, sampling precision, and the target
task. Concentration can make expectation values statistically indistinguishable
even when the formal Hilbert space dimension $d_R$ is large
\cite{DYSPGEtAl:26:arxiv,SGZ:25:arxiv,XHASCEtAl:25:arxiv}. Conversely,
carefully selected observables or encodings can preserve useful distinctions
within a much smaller effective space. Reservoir kernels describe the induced
feature geometry \cite{BB:26:arxiv}, whereas Krylov complexity and
observability describe dynamically reachable and measurable directions
\cite{IDJ:25:PRR,IJD:25:PRR}. These analyses are complementary and do not turn
formal Hilbert space dimension into a direct measure of usable capacity.

\subsection{Universality}
Universality is a property of a model family, input domain, filter class, norm,
and readout algebra. Classical reservoir theory proves universal approximation
of causal fading memory filters for specified echo state constructions
\cite{GO:18:NN}. Quantum results establish analogous guarantees for particular
dissipative systems and readouts under continuity, contraction, separation,
and algebraic closure conditions \cite{CN:19:QIP_2}. A quantum Wiener
construction provides a further guarantee within its own model class by
combining linear quantum dynamics, weak continuous measurement, and a static
nonlinear readout \cite{BB:26:arxiv}. None of these theorems implies that an
arbitrary quantum evolution and observable set is universal.

\subsection{Noise, finite sampling, and \ac{NISQ} constraints}
Conventional \ac{QRC} fixes the quantum reservoir during each readout fit and
therefore avoids gradient optimization through the quantum dynamics. This
removes one source of circuit evaluations and trainability failure, but it does
not remove calibration, parameter selection, state preparation, or measurement
cost \cite{MANBGEtAl:21:AQT,DCB:22:PRE}. Dissipation and noise can improve
performance in specific regimes by changing contraction and memory, while
excessive noise suppresses distinctions needed by the readout
\cite{CN:19:QIP_2,SASGZ:24:Qu,DCB:23:SR}. Finite sampling is a separate
limitation because small expectation differences may require many executions,
and destructive measurements can require synchronized reservoir copies or
sequence replay \cite{ATM:25:QMI,SGZ:25:arxiv,XHASCEtAl:25:arxiv}. Suitability
for \ac{NISQ} hardware should therefore be assessed through task performance
together with circuit or evolution depth, shots, measured settings,
calibration, classical postprocessing, and total quantum executions.

\section{Framework Categories and Operating Protocols}
\label{sec:qrc_framework_categories}

\Ac{QRC} is not a single architecture. Temporal memory can reside in a
retained quantum state, a measurement record with feedback, an explicit
classical window or state, or be absent from a static model. We classify an
implementation by the primary mechanism that carries information across
successive inputs because this mechanism determines state transfer and resource
cost. The categories can occur together. A retained quantum state can be combined
with a classical delay window, while a memoryless quantum feature map can feed
a recurrent classical state. We therefore record the primary memory mechanism
and any secondary mechanism rather than treating the categories as mutually
exclusive
\cite{AHZWWEtAl:24:arxiv,ATM:24:PRR,ATM:25:QMI}.

Table~\ref{tab:qrc_memory_taxonomy} summarizes the four memory architectures.
For each architecture, it identifies resources that govern the memory,
expressivity, and finite sampling properties analyzed in
Section~\ref{sec:qrc_properties_design}. After establishing these architectures,
we study composable choices that modify feature generation or training,
including multiplexing, observable expansion, nonlinear classical readout, and
outer optimization of encoding, dynamics, or measurement. This functional
taxonomy allows similar hardware to be compared under different operating
protocols
\cite{GRRO:22:NCE,AO:25:PRE,MGOEZEtAl:25:CIJNS}.

\begin{table*}[t]
\centering
\caption{Taxonomy of \acs{QRC} memory architectures. Each row identifies the
primary mechanism that carries information between successive inputs.}
\label{tab:qrc_memory_taxonomy}
\footnotesize
\setlength{\tabcolsep}{3pt}
\renewcommand{\arraystretch}{1.17}
\begin{tabularx}{\textwidth}{@{}>{\raggedright\arraybackslash}p{0.15\textwidth}>{\raggedright\arraybackslash}p{0.14\textwidth}>{\raggedright\arraybackslash}p{0.21\textwidth}>{\raggedright\arraybackslash}p{0.22\textwidth}>{\raggedright\arraybackslash}X@{}}
\toprule
\rowcolor{black!5}
\textbf{Category} &
\makecell[l]{\textbf{Temporal memory}\\\textbf{locus}} &
\makecell[l]{\textbf{Transfer between}\\\textbf{input steps}} &
\makecell[l]{\textbf{Representative}\\\textbf{realizations}} &
\makecell[l]{\textbf{Resources}\\\textbf{to report}} \\
\midrule
Intrinsic recurrent reservoir &
Retained quantum state &
The final state at step $t$ becomes the initial reservoir state at $t+1$ &
Disordered spin networks, noisy circuits, Rydberg arrays, and \acs{CV} systems
\cite{FN:17:PRA_2,CNY:20:PRA_2,BNGY:22:PQ,NAGPSEtAl:21:CP} &
Initialization and reset, input interval, sampled nodes, observables,
measurement disturbance, shots, and executions \\
\addlinespace
Measurement feedback reservoir &
Conditional quantum state and classical measurement record &
Measurement outcomes or statistics condition a later state update or control &
Weak or projective measurement protocols with conditional control
\cite{MAGSZ:23:NQI,KFY:24:PQ,FOLNM:25:PRX10} &
Instrument, outcome storage, feedback map and latency, reused trajectories or
synchronized copies, shots, and executions \\
\addlinespace
Recurrence free core with classical memory &
Classical lag window or recurrent state &
A reinitialized quantum map processes a window, or its features update a
classical state &
Windowed quantum maps and quantum features coupled to leaky or recurrent
classical updates \cite{AHZWWEtAl:24:arxiv,ATM:24:PRR,ATM:25:QMI} &
Window length or classical state dimension, classical parameters, overlapping
evaluations, data transfer, and executions \\
\addlinespace
Memoryless \acs{QELM} &
None between samples &
Every sample starts from the same reference state and is processed independently &
Fixed circuit or analog quantum feature maps
\cite{XFSACEtAl:25:QMI,MANBGEtAl:21:AQT} &
State preparation, circuit or evolution settings, observables, shots, readout
parameters, and model selection executions \\
\bottomrule
\end{tabularx}
\smallskip

\begin{minipage}{\textwidth}
\footnotesize\raggedright
\textit{Note.}
Temporal and spatial multiplexing, observable expansion, nonlinear or recurrent
readout, and outer model selection are composable modifiers rather than memory
categories. They are developed in Section~\ref{subsec:qrc_modifiers}. A
\acs{QELM} applied to a classical lag window combines the third and fourth rows.
The classical input supplies the temporal memory. The measurements, parameters,
and executions added by each modifier remain part of the reported cost.
\end{minipage}
\end{table*}
\subsection{Intrinsic recurrent reservoirs}
An intrinsic recurrent reservoir carries a quantum state from one input step
to the next. Each input modifies the retained state, fixed quantum dynamics
evolve it, and measured features feed a trained readout. The disordered spin
construction is an early example \cite{FN:17:PRA_2}. Noisy
circuits, Rydberg arrays, and \ac{CV} systems can implement the same memory
location with different channels and observables
\cite{CNY:20:PRA_2,BNGY:22:PQ,NAGPSEtAl:21:CP}. The retained state supplies
temporal dependence, while the echo state and fading memory properties depend
on the encoding, contraction rate, sampling interval, and measurement
protocol \cite{AO:25:PRE,KFS:20:SR,DCB:22:PRE}.

\subsection{Measurement feedback reservoirs}
A measurement feedback reservoir includes the measurement instrument in the
state update. Outcomes or estimated observables influence the retained state,
a reset operation, or a later control. Weak and projective protocols differ in
disturbance and statistical precision, while feedback can use quadratures,
counts, or classical statistics of earlier outcomes
\cite{MAGSZ:23:NQI,KFY:24:PQ,FOLNM:25:PRX10}. The instrument and feedback map
must be included whenever the measured system is reused to carry memory.
Observable selection or supervised measurement optimization is instead an
independent design choice when the measured outcome does not alter later
dynamics \cite{GR:26:arxiv}.

\subsection{Recurrence free quantum cores}
A recurrence free quantum core does not carry a quantum state between input
samples. Temporal context must therefore be supplied outside that core. The
windowed form adapts the classical next generation reservoir construction
\cite{GBGB:21:NC} by encoding a finite lag vector into a reinitialized quantum
map \cite{AHZWWEtAl:24:arxiv}. Other models combine memoryless quantum
features with a classical leaky or recurrent state update
\cite{ATM:24:PRR,ATM:25:QMI}. In both cases, nonlinear quantum features and
temporal memory arise in different components and their resources should be
reported separately. Independent quantum evaluations can be distributed
across processors when no quantum state links successive samples. The
achievable parallelism still depends on overlapping windows, classical state
dependencies, data transfer, and measurement cost.

\subsection{Quantum extreme learning machines}
In this survey, a \ac{QELM} is a memoryless quantum feature map. Each input is
encoded from a fixed reference state, processed by fixed quantum dynamics,
measured, and mapped to an output by a trained classical readout
\cite{XFSACEtAl:25:QMI,MANBGEtAl:21:AQT}. The quantum map carries no state
between samples. It can therefore address static data directly or temporal
data after a classical lag window has been constructed. This convention
separates memory in the model input from memory in the quantum substrate. At
increasing system size, concentration of measured features can reduce
distinguishability and increase the shots required for learning in specified
settings \cite{SGZ:25:arxiv,DYSPGEtAl:26:arxiv}. This is a feature
distinguishability and sampling limitation rather than a trainability problem
for the fixed quantum map. Some constructions reduce this concentration
within their stated assumptions through structured encodings, dynamics, or
measurements \cite{DYSPGEtAl:26:arxiv}.

\subsection{Composable feature, readout, and selection modifiers}
\label{subsec:qrc_modifiers}
Several modifiers can be applied to any memory architecture. Temporal
multiplexing samples multiple evolution times within one input interval and
produces the virtual nodes $v=1,\ldots,V$ in
Definition~\ref{def:qrc_model} \cite{FN:17:PRA_2}. Spatial multiplexing
concatenates features from independently parameterized reservoirs
\cite{NFNMK:19:PRA_2,LVRTKEtAl:26:arxiv}. Observable expansion adds local
observables, correlations, moments, or measurement channels
\cite{TN:20:arxiv,SPKRCEtAl:24:NC}. A hybrid readout adds nonlinear or recurrent
classical processing after quantum feature extraction
\cite{SSMACEtAl:25:PRA_2,NGZ:24:MLST_2,KH:25:arxiv}. Outer model selection can
choose encoding, dynamics, measurement, or modifier settings around the readout
fit. All quantum executions used for this selection belong to the model cost
\cite{KFS:20:SR,GR:26:arxiv}. These choices increase accessible features or
memory through different resources and should not be combined under one
capacity claim. Comparisons must count independent reservoir copies, evolution
samples, measurement settings, shots, and classical trainable parameters. If a
nonlinear classical readout is used, the same readout class should also be
tested with classical reservoir features.

\subsection{Operating protocols}
Operating protocols specify when the reservoir or selected subsystems are
reset, how measurements are obtained, and whether trajectories are processed
sequentially or independently. Reports should distinguish full reservoir
reset, replacement of an input subsystem, reset of measured qubits, and reset
only between sequences \cite{YSKNGEtAl:23:arxiv,MDP:23:PL,HKBAREtAl:24:NC}.
Online reuse of the same system requires an instrument model for measurement
disturbance and feedback \cite{MAGSZ:23:NQI,KFY:24:PQ,FOLNM:25:PRX10}.
Estimation with synchronized copies avoids feeding a measured copy into the
next step, but it may require parallel reservoir instances or replay of the
input prefix.
One current study reports enhanced memory under a non Markovian mechanism
\cite{SGGZ:25:arxiv}. Such persistence is not automatically beneficial because
excessive dependence on remote history or initialization can conflict with the
fading memory property. The multiplexing choices defined above apply
independently of these reset and measurement protocols.

\subsection{Reporting requirements}\par
A \ac{QRC} study should report preprocessing, encoding, reservoir dynamics,
the location of temporal memory, reset and measurement protocols, sampled
observables, shot allocation, readout class, and every optimized parameter.
For sequential experiments, it should also state whether features use reused
trajectories, synchronized copies, or replayed input prefixes. Training,
validation, and test splits must separate readout fitting from outer model
selection. Comparisons should count quantum executions, measurement settings,
classical trainable parameters, and the resources used by classical delay
embeddings or recurrent readouts. These details determine whether performance
arises from a retained quantum state, a classical window, measurement
feedback, expanded observables, or a memoryless quantum feature map. They also
determine which classical baseline and benchmark are appropriate. The full
benchmarking checklist is developed in Section~\ref{sec:qrc_benchmarking_hpc}.
\FloatBarrier

\section{Hardware Implementations and Physical Realizations}
\label{sec:qrc_hardware}

A quantum device becomes a reservoir through the controls and measurements
that turn its dynamics into task relevant features. We study hardware
realizations of \ac{QRC} by the physical variables that store the reservoir
state, the controls that encode an input, the dynamics that mix information,
and the measurements that form $\pmb{z}_{t}$. A meaningful comparison must
identify the degrees of freedom, input interface, location and lifetime of
memory, source of nonlinearity, measured observables, and cost of obtaining
those observables. We also distinguish a hardware experiment from a numerical
or theoretical model, a proposal, and a related physical reservoir study.
These distinctions are necessary because atom count, optical mode count,
qubit count, and sampled feature count describe different resources.
Table~\ref{tab:qrc-hardware-comparison} applies these comparison axes to
representative realization classes.

\begin{figure*}[t]
  \centering
  \includegraphics[width=\textwidth]{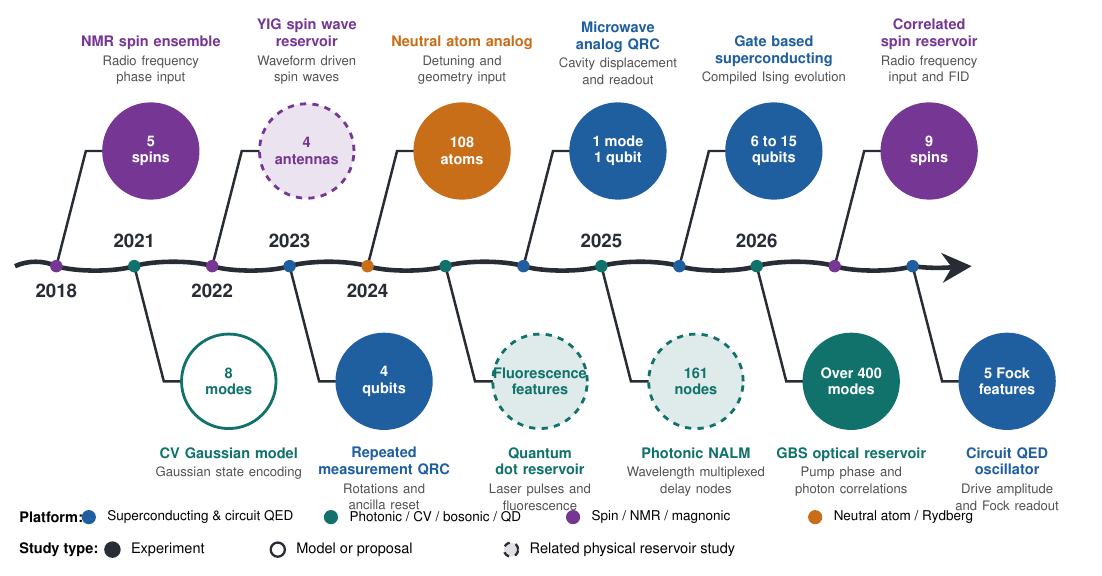}
  \caption{Timeline of representative \ac{QRC} experiments, models, and related
  physical reservoir studies from 2018 to 2026. Color denotes the platform
  family and marker style denotes the study type. Bubble text reports the
  physical scale or measured resource stated in each study. These quantities
  are heterogeneous and are not directly comparable as reservoir capacity.
  The entries cover spin, \acs{NMR}, and magnonic systems
  \cite{NMFNK:18:arxiv,NNYTHEtAl:22:arxiv,HHWXCEtAl:26:PRL_2}; photonic,
  \acs{CV}, bosonic, and quantum dot systems
  \cite{NAGPSEtAl:21:CP,T:23:NoVenue,ALFDAEtAl:25:NC,CSPMMEtAl:26:NQI};
  superconducting, microwave, and circuit QED systems
  \cite{YSKNGEtAl:23:arxiv,SPKRCEtAl:24:NC,HCLAW:25:arxiv,CDBGM:26:arxiv};
  and a neutral atom implementation \cite{AHZWWEtAl:24:arxiv}.}
  \Description{A timeline from 2018 to 2026. Each entry contains a bubble that
  states a physical scale or measured resource, a platform name, and a short
  phrase describing its input or measurement interface. Color identifies the
  platform family. Filled markers denote experiments, outlined markers denote
  models or proposals, and lightly filled dashed markers denote related
  physical reservoir studies. Bubble sizes are equal and do not encode a
  quantitative comparison.}
  \label{fig:qrc-hardware-timeline}
\end{figure*}

The platform families below follow the color groups in
Figure~\ref{fig:qrc-hardware-timeline}: spin, \ac{NMR}, and magnonic
reservoirs; photonic, \ac{CV}, bosonic, and quantum dot reservoirs;
superconducting, microwave, and circuit QED reservoirs; and neutral atom and
Rydberg reservoirs. The boundaries reflect control and measurement interfaces.
Optical bosonic and quantum dot systems are grouped by their photonic
interface, whereas microwave bosonic systems are grouped with superconducting
hardware. Figure~\ref{fig:qrc-hardware-timeline} gives a chronology and marks
the study type of each entry. It is not a ranking of platform maturity.
A 2025 conference abstract proposed reservoir enhanced sensing with trapped
ion motional modes \cite{MPMEtAl:25:DAMOP}. Because it presents an initial
concept rather than a realized or fully specified \ac{QRC} architecture, we
note it here rather than assigning a separate platform family.

\begin{table*}[t]
\caption{Comparison of representative \acs{QRC} realization classes. Labels
in the first column identify experiments (\StudyExp) and numerical or
theoretical studies, including proposals (\StudyNum). Memory and cost refer to
the cited realization, not every device in the platform family.}
\label{tab:qrc-hardware-comparison}
\centering
\footnotesize
\setlength{\tabcolsep}{2.2pt}
\renewcommand{\arraystretch}{1.17}
\begin{tabularx}{\textwidth}{@{}>{\raggedright\arraybackslash}p{0.145\textwidth}
>{\raggedright\arraybackslash}p{0.105\textwidth}
>{\raggedright\arraybackslash}p{0.105\textwidth}
>{\raggedright\arraybackslash}p{0.135\textwidth}
>{\raggedright\arraybackslash}p{0.125\textwidth}
>{\raggedright\arraybackslash}p{0.115\textwidth}
>{\raggedright\arraybackslash}X@{}}
\toprule
\rowcolor{black!5}
\makecell[l]{\textbf{Platform}\\\textbf{and studies}} &
\makecell[l]{\textbf{State}\\\textbf{DoF}} &
\textbf{Input} &
\textbf{Memory} &
\textbf{Nonlinearity} &
\textbf{Observables} &
\makecell[l]{\textbf{Dominant}\\\textbf{cost}} \\
\midrule
\makecell[l]{\textbf{Spin and \ac{NMR}}\\\textbf{registers}}\par
\StudyExp\,\cite{NMFNK:18:arxiv,HHWXCEtAl:26:PRL_2}\enspace
\StudyNum\,\cite{AKS:25:arxiv,KHSHS:24:PR,KS:25:arxiv} &
Nuclear spins &
Radio frequency pulse phase or angle &
Correlated spin state; lifetime set by dephasing and relaxation &
Input preparation, spin interactions, and observable map &
Time sampled \ac{FID} signal &
Pulse calibration, sequence replay, and signal acquisition \\
\addlinespace
\rowcolor{black!2}
\textbf{Magnonic devices}\par
\StudyExp\,\cite{NNYTHEtAl:22:arxiv,VSVMFEtAl:23:CP} &
Propagating spin waves &
Antenna drive waveform &
Wave propagation and relaxation &
Interference, dispersion, and material response &
Antenna voltages or transmission &
Bandwidth, antenna channels, and acquisition \\
\addlinespace
\textbf{\ac{CV} optical networks}\par
\StudyExp\,\cite{PHBGMEtAl:26:NP_2}\enspace
\StudyNum\,\cite{NAGPSEtAl:21:CP,BGSZ:23:PRA_2,BGSZ:24:OE} &
Optical modes and quadratures &
Displacement, squeezing, or pump phase &
Recirculating field or cavity state; loop and cavity lifetime &
Input encoding and covariance or higher order observables &
Homodyne means and correlations &
Optical loss, ensemble size, and homodyne sampling \\
\addlinespace
\rowcolor{black!2}
\makecell[l]{\textbf{Photon-counting}\\\textbf{optics}}\par
\StudyExp\,\cite{CSPMMEtAl:26:NQI,BPCCOEtAl:26:NQI,RRGSMEtAl:25:arxiv,
ZIMLSEtAl:25:SA}\enspace
\StudyNum\,\cite{NNMF:25:OQ} &
Optical modes and photon occupations &
Source, interferometer, pump, or phase controls &
Usually reset per sample; temporal context supplied by an input window &
Multiphoton interference and photon statistics &
Photon counts and correlations &
Photon budget, detector samples, and correlation estimation \\
\addlinespace
\makecell[l]{\textbf{Optical physical}\\\textbf{reservoirs}}\par
\StudyExp\,\cite{ALFDAEtAl:25:NC,TYSSNEtAl:24:AO} &
Loop pulses or excitonic populations &
Optical waveform, wavelength, or excitation pulse &
Loop delay or energy transfer lifetime &
Nonlinear transmission or excitonic energy transfer &
Intensity or fluorescence patterns &
Node timing, detector channels, and acquisition \\
\addlinespace
\rowcolor{black!2}
\makecell[l]{\textbf{Gate based}\\\textbf{superconducting}\\\textbf{circuits}}\par
\StudyExp\,\cite{SGPYY:22:SR,DHB:22:arxiv,YSKNGEtAl:23:arxiv,
HCLAW:25:arxiv,CEAN:26:CP}\enspace
\StudyNum\,\cite{DHB:22:arxiv,HCLAW:25:arxiv} &
Qubits &
Input rotations or repeated data upload &
Retained qubits, circuit history, or a classical input window &
Input encoding, entangling dynamics, and measurement probabilities &
Pauli averages or bitstring probabilities &
Circuit depth, reset, shots, calibration, and transpilation \\
\addlinespace
\makecell[l]{\textbf{Microwave and}\\\textbf{circuit QED}}\par
\StudyExp\,\cite{CDBGM:26:arxiv,SPKRCEtAl:24:NC}\enspace
\StudyNum\,\cite{DCPMGEtAl:23:NQI,DGZ:26:PRR} &
Cavity modes and transmons &
Cavity displacement or continuous drive &
Cavity state; lifetime set by decay and dephasing &
Jaynes Cummings, Kerr, or related oscillator interactions &
Quadratures, parity, or Fock populations &
Acquisition time, measurement channels, and repeated estimates \\
\addlinespace
\rowcolor{black!2}
\makecell[l]{\textbf{Neutral atom and}\\\textbf{Rydberg arrays}}\par
\StudyExp\,\cite{AHZWWEtAl:24:arxiv}\enspace
\StudyNum\,\cite{BNGY:22:PQ,LMZG:24:arxiv} &
Atomic ground and Rydberg occupations &
Detuning, Rabi drive, geometry, or an input window &
Retained atomic state in models; classical window in the reported hardware protocol &
Rydberg blockade, many body dynamics, and correlations &
Occupation bitstrings and one or two atom correlations &
Array preparation, pulse execution, geometry, and task dependent shots \\
\bottomrule
\end{tabularx}
\end{table*}

\subsection{Spin, \ac{NMR}, and Magnonic Reservoirs}

Spin reservoirs use nuclear or electronic spins as their dynamical degrees of
freedom. Inputs are commonly encoded by the phase or angle of radio frequency
pulses, after which fixed spin couplings mix the encoded state. Coherent
precession and spin interactions generate the measured features, while
dephasing and relaxation set the interval over which earlier inputs can affect
later measurements. In an \ac{NMR} experiment, a five spin molecular register
was controlled within a macroscopic ensemble and read through the ensemble
\ac{FID} signal \cite{NMFNK:18:arxiv}. Because ensemble readout is effectively
weak at the level of each molecule, repeated observations can be obtained
without the same projective reset pattern used by gate based devices. The
platform nevertheless requires explicit reporting of pulse calibration,
sampling times, relaxation scales, and the mapping from the \ac{FID} trace to
the feature vector. Later work further developed solid state nuclear spin
control and readout protocols for quantum learning tasks
\cite{NMNFNEtAl:21:NoVenue}.

A recent nine spin experiment uses radio frequency input pulses, free
induction readout, and time multiplexed samples of the measured signal
\cite{HHWXCEtAl:26:PRL_2}. Its recurrent information is associated with the
correlated spin state and its relaxation dynamics. The acquisition protocol
reinitializes the system and replays a finite input segment ending at each
prediction time, so the number of sequence replays is part of the experimental
cost. In the taxonomy of Section~\ref{sec:qrc_framework_categories}, the spin
state supplies intrinsic recurrence, while segment replay is an execution
protocol rather than an additional quantum memory. Numerical studies of
smaller spin networks separately examine distributed encoding, coupling
structure, and the relation between correlations and task performance
\cite{AKS:25:arxiv,KHSHS:24:PR,KS:25:arxiv}. These studies should not be read as
hardware demonstrations, and a spin substrate alone does not show that
entanglement is the resource responsible for a performance change.

Magnonic devices provide related physical reservoir context. In yttrium iron
garnet, a waveform excites propagating spin waves whose interference,
dispersion, and material nonlinearity transform the input. Antenna voltages or
transmitted wave amplitudes supply the classical readout, and memory is set by
propagation and relaxation times. Reported yttrium iron garnet systems address
\ac{NARMA} and dynamical prediction benchmarks as physical reservoir
experiments rather than direct \ac{QRC} demonstrations
\cite{NNYTHEtAl:22:arxiv,VSVMFEtAl:23:CP}. They remain useful comparisons for
bandwidth, multiplexing, and control cost, provided that the physical origin of
the features is not conflated with a quantum computational resource.

\subsection{Photonic, \ac{CV}, Bosonic, and Quantum Dot Reservoirs}

As reviewed in \cite{ASAPSEtAl:25:JPP}, photonic reservoirs can distribute
information across temporal, spectral, spatial, and photon number modes. The physical meaning
of the reservoir state depends on the regime. In a \ac{CV} system, quadrature
means and covariance matrices evolve under optical transformations. Input
amplitudes, displacements, squeezing parameters, or pump phases determine the
encoding, and homodyne measurements or estimated correlations form the
features. Optical recirculation and cavity lifetime can retain information
between input times. By contrast, a device that is reinitialized for every
sample supplies a memoryless photonic feature map unless a classical input
window carries the temporal context. These cases correspond to the intrinsic
recurrent and recurrence free architectures in
Section~\ref{sec:qrc_framework_categories}.

The Gaussian framework in \cite{NAGPSEtAl:21:CP} is theoretical and states
conditions for reservoir computation with Gaussian states and channels.
Numerical studies of proposed cavity loops examine how delayed recirculation,
squeezing, loss, and sampling time change fading memory
\cite{BGSZ:23:PRA_2,BGSZ:24:OE}. A later optical experiment controls memory by
shaping the pump phase of a multimode network and extracts mode selective
homodyne features \cite{PHBGMEtAl:26:NP_2}. These mechanisms are physically
different: loop delay retains an earlier optical field, whereas covariance and
photon correlations enlarge the observable set. Their resources should
therefore be stated separately.

Gaussian boson sampling provides another regime. A reported experiment uses
more than 400 optical modes and constructs features from photon count
correlations for classification \cite{CSPMMEtAl:26:NQI}. The mode count
describes the optical system, not the rank of the feature matrix or the number
of statistically resolved observables. Performance must be interpreted with
the photon budget, detector efficiency, number of samples, selected
correlations, and classical processing used to estimate those correlations.
Integrated multiphoton forecasting and a photonic learning accelerator have
been demonstrated experimentally
\cite{BPCCOEtAl:26:NQI,RRGSMEtAl:25:arxiv}, whereas photon number resolving
\ac{QRC} has been analyzed numerically \cite{NNMF:25:OQ}. These results
establish task performance under their reported protocols; an advantage claim
requires a matched classical optical or numerical baseline and full sampling
cost. A separate photonic experiment uses reservoir features for entanglement
witnessing without full state tomography \cite{ZIMLSEtAl:25:SA}.

Some optical platforms in the literature are physical reservoirs rather than
direct quantum reservoir implementations. An experimental \ac{NALM} uses
pulse circulation, interference, and nonlinear transmission to create virtual
nodes for parallel learning tasks \cite{ALFDAEtAl:25:NC}. Its memory follows
the loop delay and node timing. Semiconductor quantum dot reservoirs use
pulsed optical excitation, fluorescence, and non radiative energy transfer to
produce spatial and temporal intensity patterns \cite{T:23:NoVenue}. An
experimental device trains a linear readout on these patterns
\cite{TYSSNEtAl:24:AO}. The delayed exclusive OR
results show nonlinear temporal processing in the physical device, but do not
alone establish that coherence or another specifically quantum resource is
responsible. These systems are included because their interfaces and resource
accounting inform photonic \ac{QRC}, while their study type remains explicit.

\subsection{Superconducting, Microwave, and Circuit QED Reservoirs}

Superconducting hardware supports both gate based qubit reservoirs and analog
microwave reservoirs. In a gate based device, input dependent rotations encode
each sample, fixed entangling layers mix the qubits, and Pauli observables or
bitstring probabilities are estimated from repeated shots. Temporal memory can
reside in qubits retained between inputs, in repeated encoding within a
circuit, or in a classical input window. These cases are operationally
different even when they use the same processor. Retained qubits implement
intrinsic recurrence, whereas a reset circuit driven by an explicit input
window implements a recurrence free core in
Section~\ref{sec:qrc_framework_categories}.

Experiments on IBM superconducting processors show that device relaxation and
dephasing can alter contraction and memory, but useful noise in one task does
not make noise a general computational resource \cite{SGPYY:22:SR}. A
transmon connectivity study combines device runs with numerical simulations to
relate coupling structure to \ac{MC} \cite{DHB:22:arxiv}. In the repeated
measurement experiment of \cite{YSKNGEtAl:23:arxiv}, two system qubits retain
the reservoir state while two ancilla qubits are measured and reset after each
step. This separation makes the memory location and measurement disturbance
explicit. A multivariate forecasting experiment evaluates 6, 9, 12, and 15
qubit configurations \cite{HCLAW:25:arxiv}. Another experiment predicts
chaotic dynamics with as many as 16 qubits \cite{CEAN:26:CP}. Qubit
count alone is insufficient for comparison. Circuit depth, reset operations,
shots, calibration drift, transpilation, and any replay of the input sequence
also contribute to the implementation cost.

Circuit QED reservoirs replace or supplement a qubit register with one or more
microwave oscillator modes. A drive displaces the cavity field, while coupling
to a transmon supplies nonlinear Jaynes Cummings or related dynamics. Memory
is stored in the cavity state over its decay time and is accessed through
quadrature measurements, qubit outcomes, or reconstructed Fock populations.
Studies of coherently coupled oscillators and Jaynes Cummings reservoirs have
primarily used numerical models, including models with experimentally
calibrated parameters \cite{DCPMGEtAl:23:NQI,DGZ:26:PRR}. They establish how
coupling, loss, and sampling affect the computed features, but are distinct
from a hardware run. A direct circuit QED experiment encodes inputs in the
drive amplitude and measures the five populations $P_0$ through $P_4$ as
features \cite{CDBGM:26:arxiv}. Each reported probability is averaged over
10000 shots. The five feature channels are measured Fock populations
rather than a statement about the full oscillator dimension.

An analog microwave experiment provides another route: a continuous signal
displaces a cavity mode coupled to a qubit, and projective qubit and oscillator
parity measurement records form the reservoir features
\cite{SPKRCEtAl:24:NC}. The device processes the waveform without compiling a
universal gate sequence. Its relevant resources are cavity lifetime, drive
bandwidth, number of measurement channels, acquisition time, and the repeated
measurements used to estimate each feature. Classical microwave circuits can
obey similar nonlinear equations, so a quantum interpretation should be tied
to the prepared states and measured observables rather than to the use of
microwave hardware alone.

\subsection{Neutral Atom and Rydberg Reservoirs}

Neutral atom arrays use two level atomic states coupled through Rydberg
interactions. The atom positions set the interaction graph, while the Rabi
drive and detuning control transitions between ground and Rydberg states.
Blockade mediated many body evolution therefore supplies the reservoir map
without a compiled gate sequence. Measurements return occupation bitstrings,
from which single atom occupations and atom pair correlations can be used as
features. Numerical studies and proposals first examined how geometry,
interaction strength, and dissipation affect this construction
\cite{BNGY:22:PQ,LMZG:24:arxiv}. Retaining the atomic state between inputs
would realize intrinsic recurrence, while resetting the array for each input
window realizes the recurrence free category of
Section~\ref{sec:qrc_framework_categories}.

A direct experiment on the Aquila processor used arrays of up to 108 atoms for
forecasting and classification \cite{AHZWWEtAl:24:arxiv}. For temporal
prediction, the reported protocol encodes an explicit window of recent
classical samples and reinitializes the quantum system for each window. The
temporal memory is therefore supplied by preprocessing rather than by a
quantum state retained across prediction times. Shot count depends on the task
and protocol. The study states that its hardware runs typically use about 100
shots per data point and reports 220 measurements per embedding for one
classification task \cite{AHZWWEtAl:24:arxiv}.
The reported value of 110 shots per data point belongs to a finite sampled
simulation used in a time series comparison, not to that hardware run. Fair
comparison must include the applicable shot count together with array
preparation, pulse duration, geometry selection, and classical construction of
the input window. The count of 108 atoms describes the hardware scale, but it
does not by itself specify the number, rank, or statistical precision of the
features available to the readout.

\section{Applications}
\label{sec:qrc_applications}

The practical value of \ac{QRC} depends on whether fixed quantum dynamics can
generate useful features for a learning task without training the reservoir
itself. This role extends beyond temporal prediction to scientific surrogate
modeling, classification, quantum data inference, reconstruction, learning
under perturbations, and sequential decision making. We study these
applications by learning objective rather than domain label, and we
characterize each result by its task, implementation and resource scale,
metric, study setting, and within study baseline or limitation. This distinction
matters because the domain supplies the data and evaluation conditions but does
not define the learning problem. Financial data, for example, may support
regression, direction classification, or risk classification, each requiring a
different metric and baseline.

Table~\ref{tab:qrc_applications} groups study settings as experiments or
numerical studies; a work that reports both carries both labels. Numerical
studies include theoretical models and proposals. Within this group, simulation
denotes an abstract or idealized calculation, whereas emulation denotes a
classical solver configured for the architecture and controls of a named
hardware platform.

The applications also differ in how they represent dependence among inputs.
Recurrent reservoirs produce feature sequences for temporal tasks, whereas
static \ac{QELM} studies apply a fixed quantum feature map to individual
samples or an explicit window of classical samples. Fixing the quantum
dynamics removes quantum parameter updates from the readout fit, but the total
cost still includes preprocessing, state preparation, measurements, repeated
sequence execution, and model selection. Table~\ref{tab:qrc_applications}
summarizes selected results by task metric, implementation scale, and same
study baseline or limitation. Because the datasets, quantum
resources, and evaluation protocols differ, numerical results should not be
compared across rows.

\begin{table*}[t]
\centering
\caption{Selected \ac{QRC} application studies and their reported task
metrics, same study baselines, and principal comparison limitations. Labels
beside each citation identify experiments (\StudyExp) and numerical studies
(\StudyNum). Results use different data, preprocessing, quantum resources, and
evaluation protocols and are not directly comparable across rows.}
\label{tab:qrc_applications}
\renewcommand{\arraystretch}{1.22}
\setlength{\tabcolsep}{1.9pt}
\footnotesize
\begingroup
\hfuzz=3pt
\begin{tabularx}{0.995\textwidth}{@{}>{\raggedright\arraybackslash\hspace{0pt}}p{0.21\textwidth}
>{\raggedright\arraybackslash\hspace{0pt}}p{0.18\textwidth}
>{\raggedright\arraybackslash\hspace{0pt}}p{0.24\textwidth}
>{\raggedright\arraybackslash\hspace{0pt}}X@{}}
\toprule
\rowcolor{black!5}
\textbf{Task} &
\makecell[l]{\textbf{Implementation}\\\textbf{and resources}} &
\makecell[l]{\textbf{Metric}\\\textbf{and result}} &
\makecell[l]{\textbf{Same study baseline}\\\textbf{or limitation}} \\
\midrule
\rowcolor{black!5}
\multicolumn{4}{c}{\textbf{Temporal prediction and scientific surrogates}} \\
\addlinespace[1pt]
Lorenz 63 and \acs{ENSO} multivariate prediction\par
\StudyExp\StudyNum\,\cite{HCLAW:25:arxiv} &
Gate based model on IBM Heron R2 &
\acs{MSE} $0.0087$ and $0.0036$, respectively &
Above the tested \acs{RNN}; \acs{NVAR} and clustered \acs{ESN} are stronger in some settings \\
\acs{NARMA} and weather prediction\par
\StudyExp\,\cite{HHWXCEtAl:26:PRL_2} &
Nine spin \acs{NMR} reservoir &
Prediction error and $R^2$; lower error under the reported protocols &
Earlier \acs{QRC} and classical reservoir comparisons use unmatched physical resources \\
Mobile trajectory prediction\par
\StudyExp\StudyNum\,\cite{MCLL:23:arxiv} &
Four qubit simulation and a five qubit IBM test &
Lower \acs{MSE} than selected \acs{LSTM} and \acs{ESN} models in six simulated coordinate cases &
The hardware test uses 50 points, predicts five steps, and uses 4000 shots \\
Rayleigh B\'enard turbulence\par
\StudyNum\,\cite{PHS:23:PRR} &
Hybrid quantum and classical surrogate &
Recovery of profiles, heat transport, and low order statistics &
The target is statistical reconstruction rather than pointwise trajectory prediction \\
\addlinespace
\rowcolor{black!5}
\multicolumn{4}{c}{\textbf{Classification and quantum data inference}} \\
\addlinespace[1pt]
Earth observation image classification\par
\StudyExp\,\cite{ZSSBEEtAl:26:arxiv} &
Quantum feature extraction on IBM processors &
Best hybrid accuracy $87\%$; hardware accuracy about $86.5\%$ &
Same study ResNet50 and transfer pipelines report $83\%$ and $84\%$ under the stated split \\
Photonic entanglement witnessing\par
\StudyExp\,\cite{ZIMLSEtAl:25:SA} &
Orbital angular momentum measurement features &
Test \acs{MSE} $0.017$; $91.4\%$ of the tested entangled states correctly certified &
The shadow tomography comparison is restricted to the tested states and measurement settings \\
MNIST image classification\par
\StudyNum\,\cite{LCEGLEtAl:25:PRA_2} &
Trained autoencoder and a simulated \acs{QELM} with 5 to 12 qubits &
Test accuracy approaches $98\%$ and begins to saturate near 9 or 10 qubits &
The quantum classifier exceeds the tested full image one layer network beyond 6 qubits, but preprocessing includes a trained autoencoder \\
Credit default classification\par
\StudyExp\StudyNum\,\cite{VVVTMEtAl:25:arxiv} &
Full data device specific emulation and Aquila runs on smaller subsets &
$F_1$ is the primary metric; noiseless emulation is comparable to the tested deep network &
Hardware noise lowers $F_1$ relative to noiseless emulation \\
Biomarker outcome classification\par
\StudyExp\StudyNum\,\cite{AMVWWEtAl:26:arxiv} &
108 samples; neutral atom emulation and Aquila execution &
Emulated features give mean test accuracy comparable to classical features; hardware often reduces variation across splits &
Hardware shot limits require an aggregated feature ranking, preventing an unbiased direct comparison with the classical analysis \\
\addlinespace
\rowcolor{black!5}
\multicolumn{4}{c}{\textbf{Reconstruction and signal recovery}} \\
\addlinespace[1pt]
Image denoising\par
\StudyExp\StudyNum\,\cite{DAW:25:arxiv} &
Neutral atom emulation and a 14 atom Aquila run &
\acs{MSE}, SSIM, and Tenengrad sharpness &
Quantum features give greater sharpness and similar SSIM; the PCA baseline has lower \acs{MSE}, and hardware is weaker than emulation \\
\addlinespace
\rowcolor{black!5}
\multicolumn{4}{c}{\textbf{Robust learning and sequential decision making}} \\
\addlinespace[1pt]
Adversarial image classification\par
\StudyNum\,\cite{TTCS:25:arxiv} &
Simulated Rydberg reservoir &
Clean and robust accuracy under \acs{FGSM}, \acs{PGD}, and DeepFool &
The reported gain is limited to the tested attacks, perturbation budgets, and baseline training procedures \\
Quantum recurrent reinforcement learning\par
\StudyNum\,\cite{C:24:I22IICASSPI} &
Eight qubit \acs{QLSTM} reservoir in MiniGrid &
Average score is comparable in most tested settings &
The model trained throughout with four layers performs best in the hardest case; total training cost is not reported \\
\bottomrule
\end{tabularx}
\endgroup
\end{table*}

\subsection{Temporal Prediction and Scientific Surrogates}

Temporal prediction is a common \ac{QRC} benchmark family. Synthetic tasks
such as \ac{NARMA} and Mackey Glass test whether measured features retain
delayed inputs and expose nonlinear functions to a simple readout
\cite{FDCRM:23:SR,ATM:24:PRR,LMMB:26:PRR}. Chaotic maps and extreme event
tasks additionally test closed loop stability, valid prediction time, and
recovery of invariant statistics \cite{MHS:24:arxiv,DGZ:26:PRR}. These are
diagnostic benchmarks rather than deployed applications. Prediction error and
valid prediction time measure task performance, whereas \ac{IPC} describes a
reservoir property under a specified input distribution and readout basis
\cite{LCGZ:23:AQT}.

Performance depends on where memory is stored and how quickly it decays. In a
study of a reservoir induced by noise, an intermediate noise level improved a
specified prediction task by introducing contraction without removing all
input distinguishability \cite{FDCRM:23:SR}. This conditional result is not a
general benefit of noise. Other models retain the quantum state, encode an
explicit window of earlier samples, or combine quantum features with a
classical temporal update \cite{ATM:24:PRR,ATM:25:PRSMPES,MGOEZEtAl:25:CIJNS}.
The memory source must be stated because these designs use different quantum
execution and classical storage resources.

Observed time series include weather, climate, mobility, market, and power
demand data. Hardware studies use different mechanisms. The nine
spin experiment retains correlations in an \ac{NMR} system and replays finite
input segments, the repeated circuit observer retains selected qubits between
measurements, and the multivariate gate model constructs features on IBM
hardware \cite{HHWXCEtAl:26:PRL_2,CEAN:26:CP,HCLAW:25:arxiv}. Their sequence
lengths, resets, observables, shots, and classical baselines are not matched.
Mobile trajectory prediction provides a smaller example: a four qubit
simulation reports lower \ac{MSE} than the selected \ac{LSTM} and \ac{ESN}
configurations, while the accompanying IBM test uses only 50 points, five
prediction steps, and 4000 shots \cite{MCLL:23:arxiv}.

Financial prediction studies also represent different learning problems.
Quantum reservoirs have been simulated for food price and realized volatility
regression \cite{DGCBB:23:arxiv,LMBH:26:PRR}. A separate stock study predicts
volume and direction. Direction accuracy above $86\%$ occurs only for selected
companies and parameter settings, and some single day cases contain fewer than
30 samples. Within that study, the multilayer perceptron has the highest
direction accuracy for every company, while \ac{QRC} is stronger than selected
baselines for some error metrics \cite{OZBGE:26:arxiv}. These results should be
reported by dataset and metric rather than combined into a claim about
financial prediction.

Engineering studies remain mainly simulation based. A corrosion study uses a
simulated hybrid reservoir to predict degradation from environmental variables
\cite{TBHHSEtAl:25:arxiv}. Power demand studies evaluate exact statevector
simulation, finite shot sampling, device noise models, and quantized classical
readouts \cite{OPIS:26:arxiv,POIS:26:arxiv}. Neither study reports execution on
a quantum processor. Their results examine model behavior under measurement
and edge computing constraints, but do not demonstrate integration beside an
operating energy or monitoring system.

Scientific surrogate studies learn reduced descriptions of dynamical systems.
Simulated hybrid reservoirs have reproduced forecast trajectories or attractor
behavior for Lorenz and double scroll systems
\cite{WCGAWEtAl:24:arxiv}. For turbulent Rayleigh B\'enard flow, the targets are
profiles, heat transport, and statistical behavior rather than pointwise
reconstruction of the complete flow \cite{PHS:23:PRR}. Higher order reservoirs
have also been evaluated as nonintrusive reduced order models
\cite{JM:24:arxiv}. These studies establish task specific surrogate capability
in simulation, not a general reduction in scientific computing cost.

\subsection{Classification and Quantum Data Inference}

The quantum component in classification acts either as a temporal reservoir or
as a static feature map. A superconducting experiment processes sequential
sensor signals with a retained circuit state \cite{SGPYY:22:SR}. Static studies
reinitialize the quantum system for each image or data point. Examples include
Earth observation images, MNIST variants, and classification with Gaussian
boson sampling features
\cite{ZSSBEEtAl:26:arxiv,LCEGLEtAl:25:PRA_2,CSPMMEtAl:26:NQI}. The Earth
observation study uses 1000 training images and 200 test images and reports a
best hybrid accuracy of $87\%$, with hardware results near $86.5\%$. In the
\ac{QELM} image study, accuracy $0.976$ applies to full data amplitude encoding
with 12 simulated qubits, while performance begins to saturate near 9 or 10
qubits. These values are specific to their preprocessing and data splits.

Fixed quantum reservoirs have also been evaluated for quantum data inference.
A photonic experiment estimates an entanglement witness from selected orbital
angular momentum measurements for a restricted state family, with test
\ac{MSE} $0.017$. A negative estimated witness correctly certifies
entanglement for $91.4\%$ of the tested entangled states
\cite{ZIMLSEtAl:25:SA}. The comparison with shadow tomography applies to the
tested states and measurement configuration. A separate \ac{QELM} study
numerically estimates Werner state entanglement \cite{AAG:25:PR}. Polymer phase
classification uses simulated circuit features on microstructures generated by
self consistent field theory \cite{ISNM:26:arxiv}. Molecular activity
prediction likewise uses simulated neutral atom dynamics rather than a neutral
atom processor \cite{BAKSCEtAl:25:JCIM}. Because these studies process samples
independently, they are static quantum reservoir learning rather than recurrent
\ac{QRC}.

Risk and medical classification require separate evaluation from temporal
market forecasting. The credit default study selects $F_1$ because the data are
imbalanced and explicitly avoids accuracy as the primary metric. The full data
analysis uses noiseless emulation, whereas Aquila experiments use subsets of
approximately 1000 and 2500 samples. Noiseless reservoir features are
comparable to the tested deep network, while hardware noise reduces $F_1$
\cite{VVVTMEtAl:25:arxiv}. A medical study uses 108 biomarker samples and
compares classical features, noiseless neutral atom emulation, and Aquila
features \cite{AMVWWEtAl:26:arxiv}. Hardware features often reduce variation
across splits relative to emulation and sometimes increase mean test accuracy.
However, hardware shot limits require an aggregated feature ranking that does
not support an unbiased direct comparison with the classical analysis.
External data and repeated hardware trials are needed before either result can
be generalized to clinical practice.

\subsection{Reconstruction and Signal Recovery}

Reconstruction differs from classification because the output is a signal or
image rather than a class label. In neutral atom image denoising, a classical
decoder reconstructs clean pixels from reservoir features
\cite{DAW:25:arxiv}. The study evaluates \ac{MSE}, structural similarity, and
Tenengrad sharpness. In emulation, quantum features produce sharper images and
similar structural similarity to a PCA pipeline, while the PCA pipeline has
lower \ac{MSE}. The Aquila test uses a 14 atom chain and 200 shots per input,
compared with 1000 shots in emulation, and its reconstructions are weaker than
the emulated results. Reporting all three metrics prevents edge sharpness from
being mistaken for lower pixel error.

Signal oriented reservoirs also illustrate the distinction between a direct
quantum experiment and adjacent physical \ac{RC}. A superconducting microwave
experiment classifies continuous and modulated signals from projective
measurement trajectories \cite{SPKRCEtAl:24:NC}. It is a direct analog
\ac{QRC} experiment, although the device is small enough to simulate
classically. By contrast, optical link compensation with a laser reservoir is
a classical physical \ac{RC} experiment \cite{WYLD:25:SR}. It provides useful
deployment context for latency and physical input interfaces, but it is not a
\ac{QRC} application result.

\subsection{Robust Learning and Sequential Decision Making}

Adversarial studies ask whether a fixed quantum feature map changes classifier
sensitivity under bounded input perturbations. A simulated Rydberg reservoir
has been evaluated on MNIST family datasets under \ac{FGSM}, \ac{PGD}, and
DeepFool attacks and reports higher clean and robust accuracy than the tested
classical models \cite{TTCS:25:arxiv}. The result applies to the evaluated
attacks, perturbation budgets, preprocessing, and baseline training procedures.
It is an empirical comparison, not a general robustness guarantee.

In sequential decision tasks, the reservoir supplies a state representation
when the current observation is incomplete. A simulated \ac{QLSTM} reservoir
agent fixes the quantum recurrent parameters and trains the surrounding
classical components with asynchronous advantage actor critic
\cite{C:24:I22IICASSPI}. Average scores are comparable to models trained
throughout in most tested MiniGrid settings. In the hardest case, only the four
layer \ac{QLSTM} trained throughout reaches the best performance. Freezing the
quantum module removes updates to its circuit parameters, but the study does
not report a complete comparison of training time, quantum executions, shots,
or classical compute.

\subsection{Comparison Boundaries Across Applications}

Outcomes range from gains over the tested baselines to similar or weaker
performance. Tuned \ac{NVAR} and clustered \acp{ESN} are stronger in some
multivariate prediction settings \cite{HCLAW:25:arxiv}. A simulated \ac{QELM}
exceeds the tested full image one layer network beyond six qubits, but its
preprocessing includes a trained autoencoder and the study does not report
hardware execution \cite{LCEGLEtAl:25:PRA_2}. The hardest reinforcement
learning case favors a quantum recurrent model trained throughout
\cite{C:24:I22IICASSPI}. Hardware noise can also reduce performance relative
to noiseless emulation, as observed in credit default prediction
\cite{VVVTMEtAl:25:arxiv}.

Application claims are therefore specific to the task, data split,
preprocessing, metric, study type, and resource budget. Strong comparison
requires the same train, validation, and test partitions together with matched
classical preprocessing and model selection. It must also account for quantum
executions, measurements, sequence replay, and classical computation. The
benchmarking requirements in Section~\ref{sec:qrc_benchmarking_hpc} formalize these
conditions. Current application results do not establish a general quantum
advantage.

\section{Software, Computational Scaling, and Benchmarking}
\label{sec:qrc_benchmarking_hpc}

\Ac{QRC} experiments and simulations depend on a software stack that
determines the accessible reservoir size, dynamics model, measurement
protocol, and evaluation cost. In conventional fixed reservoir formulations,
the quantum dynamics are not updated by gradient training, so computation
shifts from quantum parameter optimization to repeated feature generation,
measurement, and classical readout fitting. Adaptive formulations add searches
over controls, observables, reservoir configurations, or readout
architectures. We study software and numerical backends, derive the resulting
quantum and classical resource costs, and specify a protocol for reproducible
comparison.

\subsection{Software and Numerical Backends}

Digital \ac{QRC} studies use general quantum programming frameworks for
circuit construction, transpilation, simulation, and hardware execution. The
early IBM processor demonstration used Qiskit and Aer
\cite{ATKWLEtAl:24:arxiv,CNY:20:PRA_2}. Aer supports several exact and noisy
simulation methods on CPUs and selected methods on GPUs
\cite{QAC:26:AER}. PennyLane provides circuit abstractions and Lightning
simulation backends~\cite{BISGAEtAl:18:arxiv,X:26:PLL}. In these load forecasting
studies, fixed reservoirs are simulated with \texttt{lightning.qubit} on an
NVIDIA A100, and an Elastic Net readout is trained outside the quantum circuit
\cite{OPIS:26:arxiv,POIS:26:arxiv}.
QuantumReservoirPy supplies injection, washout, feature extraction, and
readout interfaces over Qiskit~\cite{KMF:25:JOSS}. For Hamiltonian, bosonic,
and dissipative models, QuTiP supplies continuous time and master equation
solvers. Accelerator support depends on the selected solver and data
representation~\cite{JNN:12:CPC,LRMLHEtAl:26:PRe}.

Continuous time models can also be simulated with CUDA-Q Dynamics. Its
\texttt{evolve} interface supports closed system Schr\"odinger evolution and
Lindblad master equations, scheduled observables, and execution on multiple
\acp{GPU} through \ac{MPI}~\cite{NC:26:CUDA_Q}. A \ac{QRC} implementation must
specify the subsystem dimensions, initial state, input dependent Hamiltonian,
time grid, observables, and collapse operators. Lindblad evolution represents
unconditional dissipation. Selective measurement outcomes, conditional state
updates, and feedback require a quantum instrument or a stochastic trajectory
model. Reproducible simulations must also report the integrator, step size or
tolerance, numerical precision, Hilbert space truncation, and convergence
test.

Analog \ac{QRC} requires software support beyond pulse construction and device
submission. A review of analog quantum programming tools evaluates language
and device support together with parallelism, data and resource management,
hybrid programming, and portability~\cite{MSB:25:Qu}. Its assessment finds
that tools developed specifically for analog computation generally rely on
remote job submission and provide limited classical interaction, resource
control, and \ac{HPC} integration. For \ac{QRC}, these limitations affect
whether feature generation can be distributed, scheduled, and reproduced
across backends.

A neutral atom experiment illustrates the current software paths, although it
studies counterdiabatic optimization rather than \ac{QRC}. Control schedules
were executed on QuEra Aquila through Bloqade and Amazon Braket and on Pasqal
Orion Alpha through Pulser~\cite{ZHGCLKEtAl:26:NPUC}. These separate backend
paths provide hardware access but do not establish portable execution. The
complete code and raw device controls were not public because of provider
restrictions. Analog \ac{QRC} studies should therefore report package versions,
submitted schedule representations, translated device controls when
accessible, and the classical processing used to reconstruct the measured
features.

The reachable scale depends on the numerical representation and the reservoir
structure. Exact state vectors and dense density operators have exponential
memory cost. Distributed engines increase aggregate memory but add
communication and synchronization. Tensor network methods can reach larger
systems when entanglement and contraction width remain controlled, but they
offer no uniform gain for dense and strongly entangling reservoirs. Sparse
operators, symmetry sectors, local basis truncation, and stochastic
trajectories provide additional reductions for suitable models. Repeated
circuit topologies may also benefit from compilation reuse, parameter binding,
and batched execution. These gains depend on the backend and circuit structure
and should be measured rather than assumed
\cite{BCCCCEtAl:23:QCE,JBBB:19:SR,G:22:Qu}. Table~\ref{tab:qrc_software}
separates software roles from direct \ac{QRC} use.

\begin{table*}[t]
\centering
\caption{Software and numerical tools used in or directly relevant to
\ac{QRC} workflows. Capabilities depend on the package version, selected
backend, hardware, and numerical representation.}
\label{tab:qrc_software}
\renewcommand{\arraystretch}{1.18}
\setlength{\tabcolsep}{2.1pt}
\footnotesize
\begin{tabularx}{\textwidth}{@{}p{0.135\textwidth}p{0.205\textwidth}p{0.19\textwidth}p{0.205\textwidth}X@{}}
\toprule
\rowcolor{black!5}
\textbf{Tool} & \textbf{Role} &
\makecell[l]{\textbf{Execution}\\\textbf{support}} &
\makecell[l]{\textbf{Direct use}\\\textbf{or relevance}} &
\makecell[l]{\textbf{Required}\\\textbf{reporting}} \\
\midrule
Qiskit and Aer~\cite{ATKWLEtAl:24:arxiv,QAC:26:AER} &
Circuit construction, transpilation, noise models, simulation, and hardware
access & CPU and selected \ac{GPU} state vector, density operator, and tensor
network methods & Direct use in an IBM processor \ac{QRC}
study~\cite{CNY:20:PRA_2} & Qiskit and Aer versions, backend, transpilation
settings, simulator method, precision, and device calibration \\
PennyLane and Lightning~\cite{BISGAEtAl:18:arxiv,X:26:PLL} &
Circuit definition, fixed reservoir simulation, and optional differentiation
for adaptive variants & CPU, \ac{GPU}, tensor network, and distributed backends,
depending on the Lightning device & Direct \ac{QRC} simulation on an NVIDIA
A100~\cite{OPIS:26:arxiv,POIS:26:arxiv} & PennyLane and Lightning versions,
device name, precision, shots, differentiation method, and classical readout \\
\makecell[l]{Quantum\\ReservoirPy~\cite{KMF:25:JOSS}} &
QRC workflow for input injection, washout, feature extraction, and readout &
Inherits execution methods from its Qiskit backend & Package developed for
circuit model time series workflows & Package and Qiskit versions, reservoir
configuration, observables, washout, and readout solver \\
QuTiP~\cite{JNN:12:CPC,LRMLHEtAl:26:PRe} &
Hamiltonian, master equation, and trajectory simulation for open quantum
systems & Dense and sparse data layers, trajectories, and optional accelerator
extensions & Relevant to spin, bosonic, and dissipative \ac{QRC}; direct use
must be stated by each study & QuTiP version, solver, data layer, tolerances,
truncation, and trajectory count \\
CUDA-Q Dynamics~\cite{NC:26:CUDA_Q} &
Scheduled closed and open system evolution with observables & \ac{GPU} and
multi \ac{GPU} execution through \ac{MPI}, subject to integrator support & Used
in distributed \ac{QRC} simulation, not a deployed distributed hardware
experiment~\cite{LVRTKEtAl:26:arxiv} & CUDA-Q version, target, integrator,
time grid, dimensions, collapse operators, precision, and convergence test \\
\bottomrule
\end{tabularx}
\end{table*}

\subsection{Complete Workflow Cost and \ac{HPC}}

Fixed reservoir dynamics remove gradient updates through the quantum
evolution, but they do not remove feature generation, measurement, readout
fitting, or outer model selection. Let $T$ be the number of evaluated samples
or time indices, $G$ the number of measurement groups, $S$ the shots per group,
$R$ the number of independent protocol repetitions, and $F$ the measured
feature dimension. Here, $R$ counts repetitions of the complete protocol used
to assess variation across runs. It is distinct from synchronized state copies
required within one repetition, whose executions enter through $G$ and $S$.
Independent window circuits require a leading execution count proportional to
$T\cdot G\cdot S\cdot R$. In a replay protocol that reconstructs the input history before
measuring each time index, the number of evolved input steps is
\begin{equation}
C_{\mathrm{replay}}
= GSR\sum_{t=1}^{T}t
= GSR\frac{T(T+1)}{2},
\label{eq:qrc_replay_cost}
\end{equation}
before washout, calibration, mitigation, failed jobs, and model selection.
The quadratic sequence factor is absent when a retained state can continue
after measurement or when independent windows replace recurrence.

The parallel structure is protocol dependent. Measurement groups, shots,
independent windows, seeds, and reservoir instances can often be distributed
when hardware and scheduling permit. Time indices in a state retaining
reservoir remain causally sequential, and destructive measurements may require
sequence replay. A shot request to one \ac{QPU} is also not equivalent to an
independently scheduled \ac{HPC} task. Recurrence free formulations can expose
more parallel work, while finite sampling still sets the precision of the
estimated features~\cite{ATM:24:PRR,ATM:25:QMI}.

Classical cost can also become substantial. A feature matrix in
$\mathbb{R}^{T\times F}$ requires $O(TF)$ storage. Forming and solving the
normal equations for ridge regression costs approximately
$O(TF^{2}+F^{3})$, with other solvers changing the numerical behavior and
constants. Preprocessing, feature storage, cross validation, and searches over
encodings, observables, or reservoir parameters must therefore be counted
together with quantum execution.

Exact simulation stores $2^{N}$ complex amplitudes for an $N$ qubit state
vector and $4^{N}$ entries for a dense density operator. Under equal precision
and memory, the dense density representation reaches approximately half the
qubit count. Open system calculations may reduce this cost with sparse
operators, trajectories, tensor representations, symmetries, or local
truncations, so the representation must be reported. Partitioning a state
across nodes introduces communication that depends on operation locality and
the partition. Distributed \ac{QRC} architectures have been evaluated through
ideal and device informed simulation~\cite{LVRTKEtAl:26:arxiv}. Separate
systems work demonstrates scheduler managed access to several \acp{GPU} and
\acp{QPU} for general hybrid workloads~\cite{SRKBGEtAl:25:arxiv}. Together,
these studies describe a possible execution path, not a deployed distributed
\ac{QRC} benchmark.

\subsection{Benchmarking and Reproducibility}

Software overhead affects circuit construction, compilation, runtime, and
memory use~\cite{NSBBGEtAl:25:NCS}. It must be measured separately from task
quality. Reservoir diagnostics such as memory capacity and \ac{IPC} should
also be separated from application metrics. Baselines must match the task: an
echo state network, \ac{NVAR} model, recurrent network, or structured state
space model may be appropriate for sequence prediction, whereas classification
and reconstruction require their own domain baselines
\cite{GBGB:21:NC,WTS:25:IJPEDS}. A claim that models are matched must state the
matching variables, such as input history, accessible feature count, trainable
parameters, memory, training data, search budget, and execution cost.

\begin{table*}[t]
\centering
\caption{Minimum reporting protocol for reproducible \ac{QRC} comparisons.}
\label{tab:qrc_benchmark_protocol}
\renewcommand{\arraystretch}{1.18}
\setlength{\tabcolsep}{3.2pt}
\footnotesize
\begin{tabularx}{\textwidth}{@{}p{0.18\textwidth}X@{}}
\toprule
\rowcolor{black!5}
\textbf{Dimension} &
\makecell[l]{\textbf{Required}\\\textbf{information}} \\
\midrule
Task and data & Dataset and version; preprocessing and normalization; train,
validation, and test split; leakage controls; washout; rollout mode; primary
metric and valid prediction definition \\
Model selection and statistics & All searched encodings, dynamics,
observables, windows, readouts, and regularization values; validation
objective; total search budget and failed runs; independent seeds, splits,
device runs, and calibration periods; mean, dispersion, uncertainty interval,
and ablations \\
Quantum resources & Device and calibration time; physical and logical qubits
or modes; circuit depth and two qubit operations or analog evolution time;
state preparation and reset; measurement groups, shots, synchronized copies,
independent protocol repetitions, sequence replays, mitigation, and execution
time \\
Classical resources & Simulator representation, precision, truncation,
tolerances, and version; processor, accelerator, memory, node count, and
interconnect; feature matrix size, readout solver, training time, storage,
orchestration, and model search cost \\
Classical comparison & Task appropriate baselines; identical data and rollout;
reported matching variables; comparable tuning budget; complete results rather
than only the best run \\
\bottomrule
\end{tabularx}
\end{table*}

Task performance, reservoir diagnostics, quantum resources, classical
resources, model selection cost, and uncertainty are distinct dimensions. A
claim of advantage requires task appropriate classical baselines tuned under a
comparable search budget and the complete cost of input preparation, quantum
execution, measurement, feature storage, and readout. Shared datasets are
insufficient when preprocessing, rollout, shots, calibration, or search budgets
differ.

\section{Open Problems and Research Priorities}
\label{sec:open}

The main open problems in \ac{QRC} arise at the interfaces among encoding,
dynamics, measurement, readout, and execution resources. Their maturity
differs. Measurement design and memory control already have direct \ac{QRC}
results, configured multitask learning has one focused demonstration, and
federated learning and attention are motivated largely by adjacent fields. We
therefore state what is known for each topic, identify the missing result, and
specify a test that could resolve it. None of these directions presumes a
general quantum advantage.

\subsection{Measurement Design under Finite Sampling}

Measurement design has three distinct components: selecting implementable
observables, estimating them under a finite shot budget, and modeling how
measurement changes later reservoir dynamics. Kernel methods can optimize an
observable for a fixed reservoir and training set~\cite{GR:26:arxiv}, while
local observable and classical shadow constructions constrain estimation
cost~\cite{ACAW:26:arxiv}. Postmeasurement denoising addresses a separate
feature estimation problem~\cite{ATM:25:QMI}. Weak, projective, and midcircuit
measurements can additionally modify the state carried to the next input
\cite{MAGSZ:23:NQI}.

The remaining optimization problem is to minimize validation error subject to
locality, basis rotation, measurement grouping, shot allocation, state replay,
and feedback constraints. Random matrix observables have been tested on five
atom and five qubit reservoirs, so their scaling remains to be established
\cite{TFHS:25:MLST_2}. Midcircuit measurement with feedforward has structured
complexity results and a reservoir style benchmark, but this does not establish
a general supervised \ac{QRC} advantage~\cite{CE:26:PRL_2}. A decisive hardware
study should test whether an observable selected on validation data remains
useful under calibration drift and whether its gain survives the cost of
estimating it.

\subsection{Memory Engineering and Stability}

\Ac{QRC} memory depends on a balance between retention and contraction.
Dissipative models can satisfy fading memory, reset rates can tune retained
history, and numerical studies show echo state behavior and extended input
influence under specified noisy or non Markovian models
\cite{SASGZ:24:Qu,MDP:23:PL,MRNP:24:V_2IICQCEQ,SGGZ:25:arxiv}. A hardware
protocol based on midcircuit measurement and deterministic reset sustains
temporal inference beyond a single coherence interval
\cite{HKBAREtAl:24:NC}. A photonic quantum memristor has also been demonstrated
experimentally, whereas its \ac{QRC} application was evaluated numerically
\cite{SMPAMEtAl:22:NPh}.

Reset, environmental dissipation, observed measurement, and hidden
environmental memory produce different channels and resource costs. A unified
theory should relate their contraction rates, input distinguishability, task
memory capacity, finite sampling behavior, and stability. Hardware tests
should compare these mechanisms on the same task with the same input history,
measurement budget, and total execution cost.

\subsection{Multitask Capacity of Configured \ac{QRC}}

Configured \ac{QRC} has been evaluated with one reservoir optimized through a
joint multitask objective and separate linear readouts for the individual
tasks~\cite{XZQCZEtAl:23:SB}. This demonstrates feasibility for that model, not
a general multitask capacity. Such a capacity should depend on the rank and
conditioning of the accessible feature matrix, overlap among observables
relevant to each task, finite training data, measurement allocation, and the
cost of configuring the reservoir. A useful experiment should compare one
shared reservoir with separately optimized reservoirs under the same total
quantum execution and search budget, and should report the increase in each
task loss caused by joint training. Continual learning is a separate problem
that also requires sequential task arrival and a measure of forgetting.

\subsection{Federated and Distributed \ac{QRC}}

Federated \ac{QRC}, distributed execution of one \ac{QRC} model, and learning
from partitioned quantum states are different problems. In a federated
setting, clients retain local classical data and reservoirs and communicate
readout weights or sufficient statistics. This requires aligned feature
definitions and an explicit privacy model~\cite{BCA:25:QMI}. Distributed
execution instead partitions a reservoir or feature map across devices.
Current distributed \ac{QRC} results are based on ideal and device informed
simulation~\cite{LVRTKEtAl:26:arxiv}. Learning functions of partitioned
quantum states imposes a third set of measurement and communication
constraints~\cite{GFAGZ:26:arxiv}.

Fixed reservoirs do not provide privacy automatically, and statistics such as
$\pmb{Z}_{\mathrm{tr}}\pmb{Z}_{\mathrm{tr}}^{\mathsf{T}}$ and
$\pmb{Y}_{\mathrm{tr}}\pmb{Z}_{\mathrm{tr}}^{\mathsf{T}}$ can reveal
information about local features.
Classical federated reservoir studies also show sensitivity to client and data
heterogeneity~\cite{LNSSHEtAl:25:arxiv}. A credible federated \ac{QRC}
protocol must specify the threat model, secure aggregation or privacy
mechanism, client data heterogeneity, feature alignment, calibration mismatch,
shot allocation, communication volume, and performance relative to local and
centralized baselines.

\subsection{Attention over Quantum Reservoir Features}

Attention has been combined with selected classical reservoir and frozen layer
models~\cite{LMDWSEtAl:23:P3AICM,SKOU:24:PRA_2,SBMKAEtAl:20:arxiv}.
Task comparisons on collapse prediction and chaotic extrapolation show
model dependent behavior, but they do not isolate the attention map as the
cause~\cite{ZGL:26:arxiv}. Quantum circuits have also been used inside the
attention calculation. One construction encodes token, positional, query, and
key representations in parametrized quantum circuits and obtains each query
and key dot product from an ancilla measurement. The value map and the
remaining decoder operations are classical~\cite{SSKFCEtAl:25:JCTC}. In molecular
generation experiments simulated with CUDA-Q, the model used six active qubits
and performed comparably to matched classical models. This is a quantum
attention model rather than \ac{QRC}: its circuits are trained to form
embeddings and attention scores, not fixed dynamics that produce reservoir
features. The construction also leaves the quadratic dependence on sequence
length unchanged.

These studies do not demonstrate attention over \ac{QRC} measurements. In
a \ac{QRC} workflow, attention would normally act on a stored classical history
of observables, virtual nodes, or learned summaries. It therefore adds
classical memory, trainable parameters, and context dependent computation.

A decisive study should compare the same quantum features with linear,
recurrent, and attention readouts. It should also compare the attention model
with the same architecture trained on matched classical reservoir features and
vary the context length, feature count, parameter count, and measurement
budget. The question is whether quantum generated features improve a matched
attention model after these resources are counted.

\subsection{Standards for \ac{QRC} Advantage Claims}

Claims of \ac{QRC} advantage should be separated at three levels. An empirical
improvement is a statistically supported gain over matched baselines under one
fixed protocol. A scaling result shows that the gain persists as the data,
reservoir size, target precision, and noise vary under complete resource
accounting. A formal advantage additionally specifies a task family, classical
comparison class, computational assumption, error criterion, and efficient
quantum measurement procedure
\cite{BAS:24:arxiv,CC:25:arxiv,LFPEG:25:arxiv}.

Current \ac{QRC} studies provide task specific empirical results and partial
theory, but not this complete chain. Concentration can reduce accessible
feature distinguishability, and recent \ac{NARMA} studies show that results
depend strongly on benchmark and protocol choices
\cite{DYSPGEtAl:26:arxiv,KSPBDEtAl:25:arxiv,SGZ:25:arxiv}. Comparisons of
variational quantum and classical time series models reach a similarly
cautious conclusion under full resource accounting~\cite{FKTH:26:MLST_2}.

These standards need a concrete set of tasks before they can be applied, and
CogScale can supply that benchmark suite for evaluating a \ac{QRC} architecture
against tuned classical baselines. It contains 14
scalable synthetic sequence tasks and compares seven classical architectures,
including an \ac{ESN}, under trainable parameter budgets of 1000, 10000, and
100000 and at two reported difficulty levels~\cite{BDH:26:arxiv}. Its results
separate basic retention, where classical recurrent models and \acp{ESN} are
competitive at small budgets, from reasoning and manipulation tasks that favor
attention and modern state space models as difficulty increases. CogScale
treats the \ac{ESN} separately from trained models by searching the leaking
rate, spectral radius, and input scaling, then selecting the ridge penalty on
validation data. A \ac{QRC} architecture can then be evaluated on the same task generators,
seeds, and metrics against these matched classical baselines. Trainable
parameter matching alone remains insufficient, however: a quantum extension
must also match reservoir selection effort and report the physical reservoir
size, measured feature count, input sequence executions, shots, and readout
cost.

The open problem is to connect a plausible source of quantum hardness to
features that can be estimated efficiently and to a complete workflow benefit
after input preparation, measurement, readout, and model selection. The
reporting protocol in Section~\ref{sec:qrc_benchmarking_hpc} supplies the
minimum standard for testing such claims.

\section{Conclusion}
\label{sec:conclusion}

\Ac{QRC} generates temporal features through fixed or selectively configured
quantum dynamics and concentrates learning in a classical readout. We organized
the field around a common system model that links preprocessing, encoding,
reservoir dynamics, measurement and observable extraction, readout, and
execution resources. From this model, we derived a taxonomy that classifies
architectures by their primary memory mechanism while recording encoding,
measurement, feedback, and readout choices separately. This separation allows
systems that share hardware but differ in state transfer and measurement cost
to be compared by computational role rather than platform label or Hilbert
space dimension. Neither of these determines the features available to the readout,
which instead follow from the encoding, dynamics, and measurement.

Across these stages, the reported results point to the same conclusion. On spin,
photonic, superconducting, bosonic, and neutral atom platforms, hardware
experiments establish task specific feasibility while exposing the reset,
sampling, calibration, and sequence replay costs that physical scale alone
conceals. Application results range from gains over selected baselines to
comparable or weaker performance, and they depend on data splits, preprocessing,
metrics, noise models, and resource budgets that are rarely matched across studies. The
software and scaling analysis makes the associated execution, storage,
fitting, and reporting costs explicit. Current results therefore do
not establish a broad quantum advantage over well matched classical reservoirs.

Closing this gap requires standards as well as hardware advances. We
distinguish advantage claims at three levels: empirical, scaling, and formal.
Each level requires matched baselines, complete resource accounting,
uncertainty and shot budget reporting, reproducible software and measurement
protocols, and theory connecting accessible observables to memory,
expressivity, stability, and trainability under noise. The reporting protocol
specifies the minimum information needed for such claims. Applied consistently,
it lets future work in memory control, measurement design, scalable hardware,
and efficient feature extraction be judged on operational resources and
validation level rather than on nominal state space size.

\bibliographystyle{ACM-Reference-Format}
\bibliography{QRC_refs}
\appendix

\end{document}